\documentclass[a4paper,fleqn,usenatbib]{mnras}
\usepackage{graphicx,natbib}
\usepackage{psfrag}
\usepackage{amssymb}
\usepackage[dvips]{color}
\usepackage{tikz}  
\usepackage{hyperref}
\usepackage{amsfonts}
\usepackage{listings}
\usepackage{caption}

\title[Embedded clusters in Carina complex]{Near-infrared study of new embedded clusters in the Carina complex}

\author[R. A. P. Oliveira, E. Bica and C. Bonatto]{R. A. P. Oliveira$^{1}$\thanks{Present address: IAG - Universidade de S\~ao Paulo, Brazil},  E. Bica$^1$, and C. Bonatto$^1$\\
$^1$ Departamento de Astronomia, Universidade Federal do Rio Grande do Sul, 
Av. Bento Gon\c{c}alves 9500\\
Porto Alegre 91501-970, RS, Brazil}
\date{Accepted 2018 January 27. Received 2018 January 20; in original form 2017 August 19}

\begin{document}

\pagerange{\pageref{firstpage}--\pageref{lastpage}}

\maketitle

\label{firstpage}

\begin{abstract} 

We analyse the nature of a sample of stellar overdensities that we found projected on the Carina complex. This study is based on 2MASS photometry and involves the photometry decontamination of field stars, elaboration of intrinsic colour-magnitude diagrams $J\times(J-K_s)$, colour-colour diagrams $(J-H)\times(H-K_s)$ and radial density profiles, in order to determine the structure and the main astrophysical parameters of the best candidates. The verification of an overdensity as an embedded cluster requires a CMD consistent with a PMS content and MS stars, if any. From these results, we are able to verify if they are, in fact, embedded clusters. The results were, in general, rewarding: in a sample of 101 overdensities, the analysis provided 15 candidates, of which three were previously catalogued as clusters (CCCP-Cl\,16, Treasure Chest and FSR\,1555), and the 12 remaining are discoveries that provided significant results, with ages not above 4.5 Myr and distances compatible with the studied complex. The resulting values for the differential reddening of most candidates were relatively high, confirming that these clusters are still (partially or fully) embedded in the surrounding gas and dust, as a rule within a shell. Histograms with the distribution of the masses, ages and distances were also produced, to give an overview of the results. We conclude that all the 12 newly found embedded clusters are related to the Carina complex.
\end{abstract}



\section{Introduction}
\label{sec:1}

Stellar clusters are characterized as a group of stars gravitationally bound and sharing a common origin: they were formed from the same cloud of gas and dust and, therefore, have basically the same age, distance and initial chemical composition. Exceptions are globular clusters with multiple stellar populations (\citealp{Piotto15}). 

The embedded clusters (ECs), precursors of the open clusters (OCs), are very young structures, still in formation stage and partially or completely embedded in their parental cloud \citep{Lada03}. For this reason, they can hardly be studied in optical wavelengths; but infrared studies and instrumentations have changed this scenario, since infrared radiation is less affected by interstellar dust extinction.

The great majority of the stars does not form isolated in the interstellar medium but, instead, in embedded clusters (70$-$90$\%$, according to \citealp{Lada03}), inside the densest cores of the giant molecular clouds (GMCs). In particular, stars form through the gravitational collapse of high density fluctuations in the GMC, creating a non-uniform structure with different densities. Then this structure becomes gravitationally unstable, fragments and forms protostellar seeds, where the embedded cluster is formed by gas accretion \citep{Clarke00} and subsequent star formation takes place \citep{Adams06}.

Fragmentation and collapse are well explained by the Jeans instability theory, when the internal gas pressure is not strong enough to counterbalance the gravitational pressure or, in other words, when its mass exceeds the Jeans limit ($M_J \propto T^{3/2}\rho^{-1/2}$).

The study of ECs involves the formation and initial evolution of star clusters, stars and planetary systems; their interaction with the surrounding stellar material and dynamical evolution, producing field stars and, in part, ending up as OCs. After formation, the ECs interact strongly with their progenitor molecular cloud, disrupting the surrounding gaseous environment by injecting energy and momentum through some mechanisms (e.g., protostellar outflows, photoionization, stellar winds and radiation pressure, see \citealp{Morales13}). These mechanisms generate pressure gradients in the molecular cloud and expel the remaining gas and dust, creating a shell (or bubble, studied in mid-infrared wavelengths, e.g. \citealp{Churchwell06, Watson08}) structure, where the cluster finally becomes transparent at optical wavelengths. Star formation ceases when all the parent gas is dispersed. The embedded phase of cluster evolution typically lasts less than 5 Myr.

Very few ECs (4$-$7$\%$, \citealp{Lada03}) survive the emergence of the molecular cloud, because much of the mass has been abruptly lost, i.e., the amount of energy released by all massive stars exceeds the total binding energy of the cluster. Hence, the fate of an EC is determined by the mass of its massive stars, the star-formation efficiency [SFE = M$_\star$/(M$_\star$+M$_{gas}$)] and the timescale for gas dispersion. The ejection of residual gas may be very destructive, so that surviving clusters can lose more than 50$\%$ of their stars, a process called infant weight-loss \citep{Tutukov78}.

A search for star clusters consists in the detection of an excess density of stars compared to the background, i.e., an overdensity. Unlike open and globular clusters, the embedded clusters are not found isolated. Instead, usually, the molecular cloud complexes enclose multiple ECs and distributed populations of isolated stars \citep{Allen07, Camargo15}. In the present study, the search for overdensities is performed in the Carina complex, a region of intense and ongoing active star formation, which contains nebulae, several embedded and open clusters, filaments, pillars \citep{Smith05} and dust cavities, suggesting a turbulent history. Finally, we verify if the candidate overdensities are star clusters.

This paper is organized as follows. In Sect.~\ref{sec:2} we present the data and detail the methods and applied tools. In Sect.~\ref{sec:3} we exhibit the overdensity sample, and discuss the achieved results for fundamental parameters for the best candidates. Finally, in Sect.~\ref{sec:4}, we provide the concluding remarks.


\section{Data and Methods}
\label{sec:2}

This analysis is based on the near-infrared photometry of an overdensity sample and in probing them for structure and photometric parameters. The photometric data of the 2MASS All-Sky Catalog of Point Sources \citep{Cutri03}, centered at the Trumpler 16 coordinates ($\alpha_{J2000} = 10^h45^m10.0^s$, $\delta_{J2000} = -59^{\circ}43'00"$), were downloaded from VizieR\footnote{VizieR catalogue access tool is maintained by the Centre de Donn\'ees astronomiques de Strasbourg (CDS), France.}, with a radius of 90 arcmin and limited to 0.1 mag for the uncertainties in magnitude of each band $J$ (1.253$\,\mu$m), $H$ (1.662$\,\mu$m) and $K_{s}$ (2.159$\,\mu$m). The routines explained below were developed by one of us (C. B.).

\vspace{-0.1cm}\subsection{Search for overdensities}
\label{sec:2.1}

Stellar overdensities are detected with an algorithm that produces density maps and scans them, searching for regions with a higher density of stars as compared to the surroundings. Since overdensities may show up in different scales, the density maps are built with several spatial resolutions: $0\arcmin.2\times 0\arcmin.2$, $0\arcmin.33\times 0\arcmin.33$, $0\arcmin.5\times 0\arcmin.5$, $1\arcmin\times 1\arcmin$, $2\arcmin.5\times 2\arcmin.5$ and $5\arcmin\times 5\arcmin$. As an additional constraint we require that each overdensity has three    as the minimum number of stars to be considered as a candidate.

In summary, the algorithm first finds the point with the maximum density of stars; then it computes the density in neighbouring rings, stopping when the density becomes equivalent to the Poisson fluctuation. This gives the overdensity size and the floor level. Consequently, this procedure considers a local background for each overdensity. Subtraction of the floor density gives the number of member stars, which is then compared to the minimum number constraint. We also require that the overdensities must have amplitudes at least 3-sigma above the background fluctuation.

To quantify how significant an overdensity is and to certify that it is not a field fluctuation, the algorithm inserts, for each overdensity radius, equal rings in thousands of random positions and counts how many stars are found in each ring, thus giving rise to a probability density distribution (PDD) of field fluctuations. The farther the overdensity count is from the field PDD, the greater is the probability of this overdensity to be real. In this paper, we set 75\% as the minimum probability for the candidates.

\subsection{Field-star decontamination}
\label{sec:2.2}

To uncover the intrinsic morphology of CMDs, we employ the statistical algorithm for field-star decontamination described in \citet{Bonatto07} to the original photometry, widely used since then (e.g. \citealp{Camargo16}). The present version divides the CMD into a 2D grid, with axes $J$ and $(J-K_s)$, measures the relative number densities of cluster candidate and field stars and, then, subtracts the expected number of field stars in each cell.

\subsection{Fundamental parameters}
\label{sec:2.3}
The fundamental parameters are derived from the decontamined photometry essentially following the method outlined in \citet{Lima14}, \citet{Bonatto12a} and \citet{Bonatto12b}, now using the upgraded isochrone set of PARSEC \citep{Bressan12}. The output parameters are total cluster mass $M_{tot}$ (with a mass extrapolation $M_{ext}$ for all cluster members), age in $10^6$ years, apparent distance modulus (providing the distance $d_{Sun}$), colour excess $E(J-K_s)$, and differential reddening, described as a gaussian centered at $\overline{A_V}(DR)$ with $\sigma_{DR}$ as dispersion.

\subsection{Structural analysis}
\label{sec:2.4}

The structure of each cluster is determined by its stellar radial density profile (RDP). First, it is applied a colour-magnitude filter in the CMD, to isolate the stars with high probability of being cluster members, creating an intrinsic stellar RDP and enhancing the contrast relative to the background \citep{Bonatto10, Bonatto11}. The RDP is built from star counts within concentric rings, divided by the respective area, providing the surface stellar density (arcmin$^{-2}$). The King profile does not necessarily apply to ECs, because they did not reach dynamical relaxation yet, since they are too young. For more details, see Sect.~\ref{sec:3.2}.


\section{Results}
\label{sec:3}

\begin{figure}
\vspace{3mm}
\centering
\resizebox{0.475\textwidth}{!}{\includegraphics{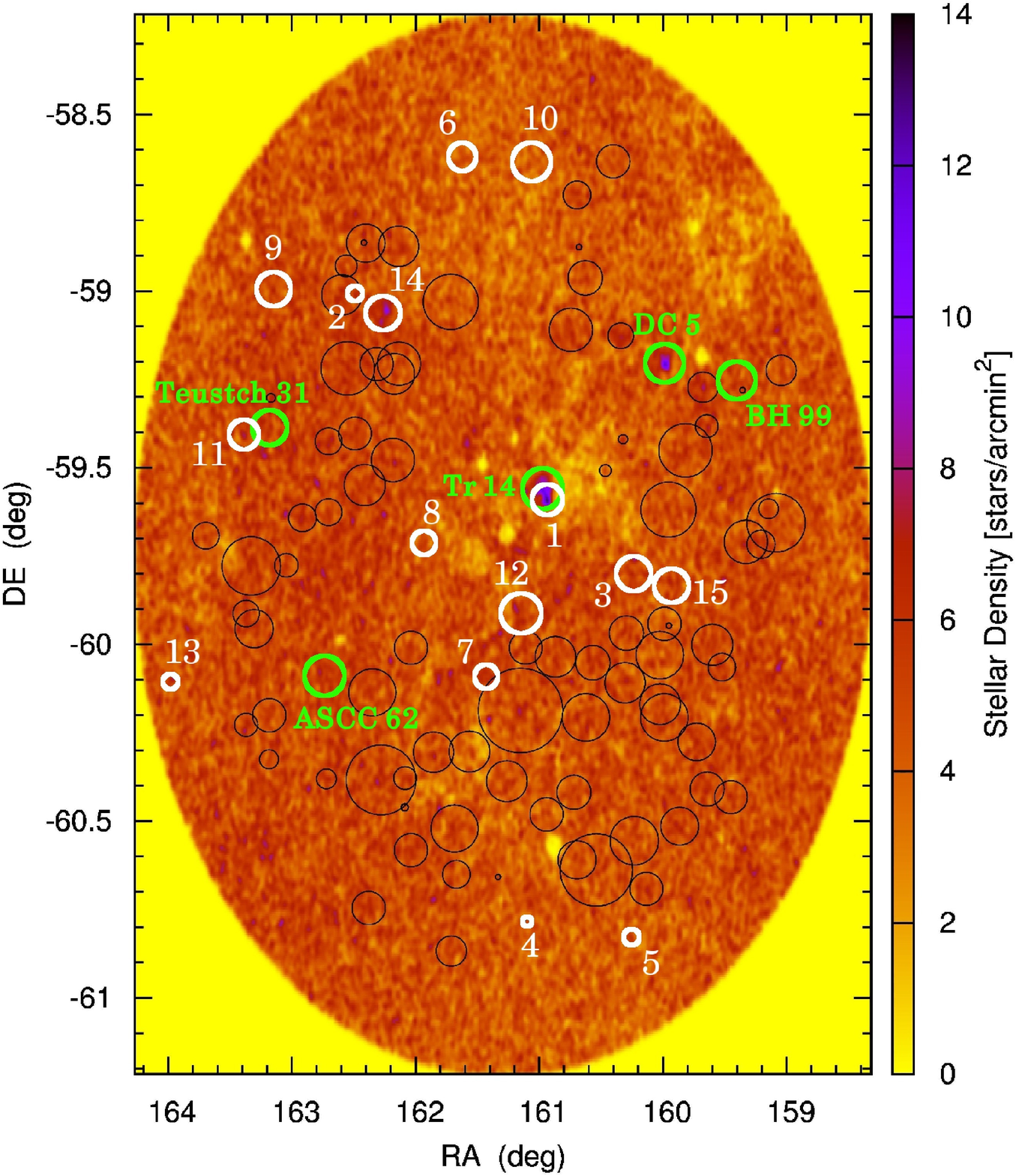}}
\caption[]{The 101 high-probability overdensities detected by the algorithm are shown with its respective radius. Some known clusters in the area are highlighted with green circles, along with the 15 candidates studied here (white circles) and field fluctuations.}
\label{f1}
\end{figure}

The search for overdensities provided 101 high-probability overdensities (probability $>50\%$, shown in Fig.~\ref{f1}). All of them were analysed separately and 15 were selected as being promising candidates for ECs, named as ``Oliveira\,1" to ``Oliveira\,15". Their equatorial coordinates, radius and size of comparison field are shown in Table~\ref{tab1}. Well-known clusters, such as Trumpler\,14 and Teutsch\,31 \citep{Dias02}, were detected and are indicated in Fig.~\ref{f1}.

\begin{figure}
\centering
\resizebox{0.46\textwidth}{!}{\includegraphics{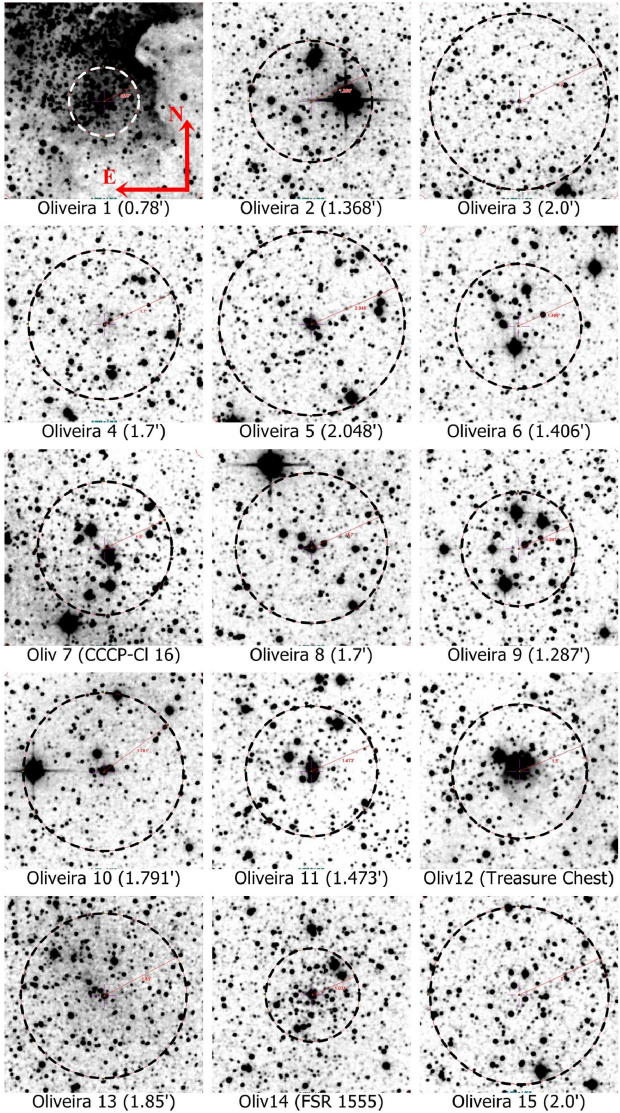}}
\caption[]{2MASS (H band) images of the 15 candidates, with each radius highlighted, all in same scale ($4\arcmin.5\times 4\arcmin.5$). The radii and comparison fields are shown in Table~\ref{tab1}.}
\label{f2}
\end{figure}

Three of the 15 candidates were previously catalogued: $(i)$ Oliveira\,7 is CCCP-Cl\,16 \citep{Feigelson11}; $(ii)$ Oliveira\,12 was included in \citet{Dutra01} and referred to as Treasure Chest \citep{Smith05}; and $(iii)$ Oliveira\,14 is FSR\,1555 \citep{Froebrich07}. FSR\,1555 has an age of $1.5\,\pm\,0.5$\,Gyr \citep{Bonatto09}, so the CMD program were not applied, because the present version has an upper limit age of 100 Myr. Fig.~\ref{f2} shows the 15 best candidates, in the H band of the near-infrared 2MASS, all in same scale ($4\arcmin.5\times 4\arcmin.5$). Additionally, Fig.~\ref{f3} presents four of them, in false-coloured images: Oliveira\,1 (2MASS-JHK), Oliveira\,4 (WISE\footnote{NASA's Wide-field Infrared Survey Explorer (mid-infrared).}), Oliveira\,10 (DSS\footnote{Digitized Sky Survey, a digital set of all-sky photographic surveys, performed with Palomar and UK Schmidt telescopes.}) and Oliveira\,13 (Herschel\footnote{ESA's Herschel Space Observatory (far-infrared).}), in order to enhance particular stellar or nebular features.

\subsection{Colour-magnitude and colour-colour diagrams}
\label{sec:3.1}

After field-star decontamination, the CMDs program was applied for each candidate. The parameters were limited to: $10-1\,000\,$M$_\odot$ for cluster mass, $0-5\,$Myr for age, $9-13\,$mag for apparent distance modulus (distance of 0.63 to 4$\,$kpc) and $0-2\,$mag for colour excess. The upper limits for $\overline{A_V}(DR)$ and $\sigma_{DR}$ were selected as low, to a better adjustment of the isochrones and simulated parameters in the Hess model. Furthermore, it was selected a SFS assuming star formation over the full cluster lifetime, with a SFR linearly decaying over time, and a Kroupa IMF \citep{Kroupa01}.

In summary, the algorithm first builds the Hess diagram of the observed CMD, and then it builds simulated Hess diagrams by varying parameters such as distance modulus, colour excess, age and star-formation spread, stellar mass and differential reddening, to emulate the observed one according to \citet{Bonatto12a}. Parameters corresponding to the simulation that minimizes the residuals with respect to the observed Hess diagram are taken as representative of the cluster. The method uses PARSEC isochrones that include MS and their extensions to PMS.

As for the differential reddening (DR), for simplicity we assume that DR across the cluster can be described by a Normal distribution function characterized by the mean value $\overline{A_V}(DR)$ and dispersion $\sigma_{DR}$. This implies that any star, regardless of its location in the cluster, can have a DR value $A_V(DR)$ with a probability proportional to $\exp[-(A_V(DR)-\overline{A_V}(DR))^2/2\sigma_{DR}^2]$. In practice, it produces a spreading of the Hess densities along the reddening vector.

\begin{table}
\centering
{\footnotesize
\caption{General data of the cluster candidates.}
\renewcommand{\tabcolsep}{1.0mm}
\renewcommand{\arraystretch}{1,2}
\begin{tabular}{lrrrrr}
\hline
\hline
Target&$\alpha(2000)$&$\delta(2000)$&Radius&Field\\
&(h\,m\,s)&$(^{\circ}\,^{\prime}\,^{\prime\prime})$&$(arcmin)$&$(arcmin)$ \\
($1$)&($2$)&($3$)&($4$)&($5$)\\
\hline
Oliveira\,1 &10:43:41.2&-59:35:50&0.78&1.1$\,-\,$1.5\\
Oliveira\,2 &10:50:10.5&-58:57:45&1.368&2.0$\,-\,$6.8\\
Oliveira\,3 &10:40:56.5&-59:48:00&2.0&3.0$\,-\,$7.0\\
Oliveira\,4 &10:44:38.5&-60:48:03&1.7&3.0$\,-\,$6.9\\
Oliveira\,5 &10:40:58.9&-60:49:48&2.048&4.1$\,-\,$8.4\\
Oliveira\,6 &10:47:03.0&-58:35:26&1.406&1.9$\,-\,$4.5\\
Oliveira\,7\textsuperscript{$\dagger$} &10:45:42.2&-60:05:27&1.5&2.6$\,-\,$7.5\\
Oliveira\,8 &10:47:53.1&-59:43:14&1.7&2.5$\,-\,$5.7\\
Oliveira\,9 &10:51:52.4&-58:59:24&1.287&3.2$\,-\,$6.8\\
Oliveira\,10 &10:46:11.5&-58:39:13&1.791&2.9$\,-\,$4.5\\
Oliveira\,11 &10:53:47.2&-59:22:50&1.473&3.1$\,-\,$6.3\\
Oliveira\,12\textsuperscript{$\dagger$} &10:45:53.6&-59:57:03&1.5&2.5$\,-\,$6.5\\
Oliveira\,13 &10:56:29.7&-60:06:11& 1.85&4.3$\,-\,$7.9\\
Oliveira\,14\textsuperscript{$\dagger$} &10:48:56.2&-59:02:52&1.037&1.0$\,-\,$3.5\\
Oliveira\,15 &10:39:23.0&-59:45:13&2.0&3.2$\,-\,$7.0\\

\hline
\end{tabular}
\begin{list}{Table Notes.}
\item Col. $1$: designations of the 15 candidates.  Cols. $2-3$: Central equatorial coordinates. Col. $4$: Cluster radius. Col $5$: Inner and outer radius of the comparison field (annulus shape). $\dagger$: The cluster is catalogued in the literature (see text below). 
\end{list}
\label{tab1}
}
\end{table}

\begin{figure}
\centering
\resizebox{0.47\textwidth}{!}{\includegraphics{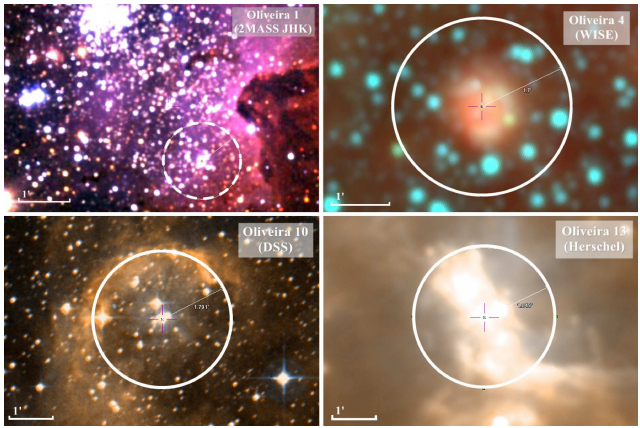}}
\caption[]{RGB composite images of four candidates in different systems (2MASS, WISE, DSS and Herschel), enhancing particular stellar and nebular features.}
\label{f3}
\end{figure}

\begin{figure*}
\resizebox{0.41\textwidth}{!}{\includegraphics{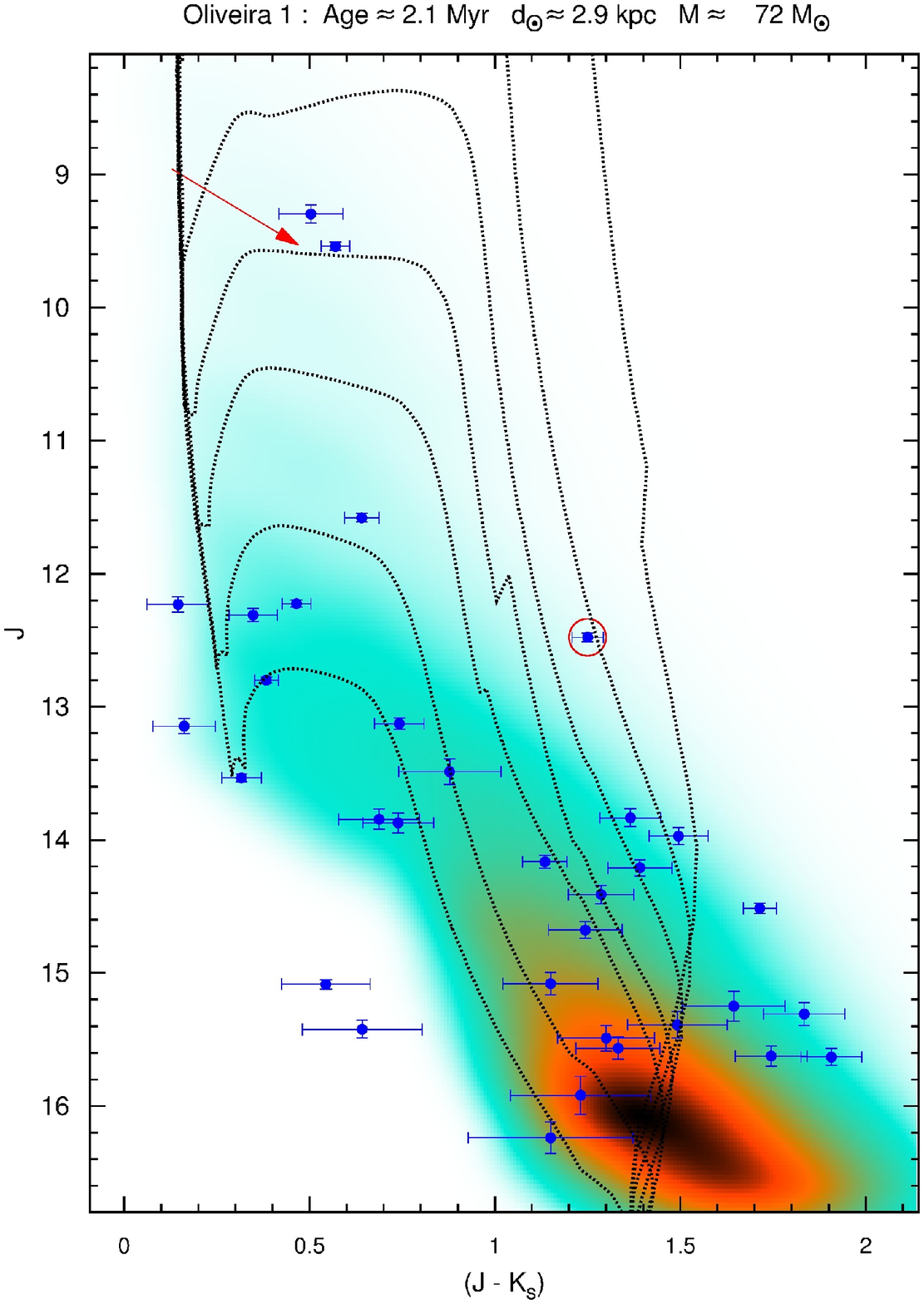}}
\hspace{0.5cm}\resizebox{0.41\textwidth}{!}{\includegraphics{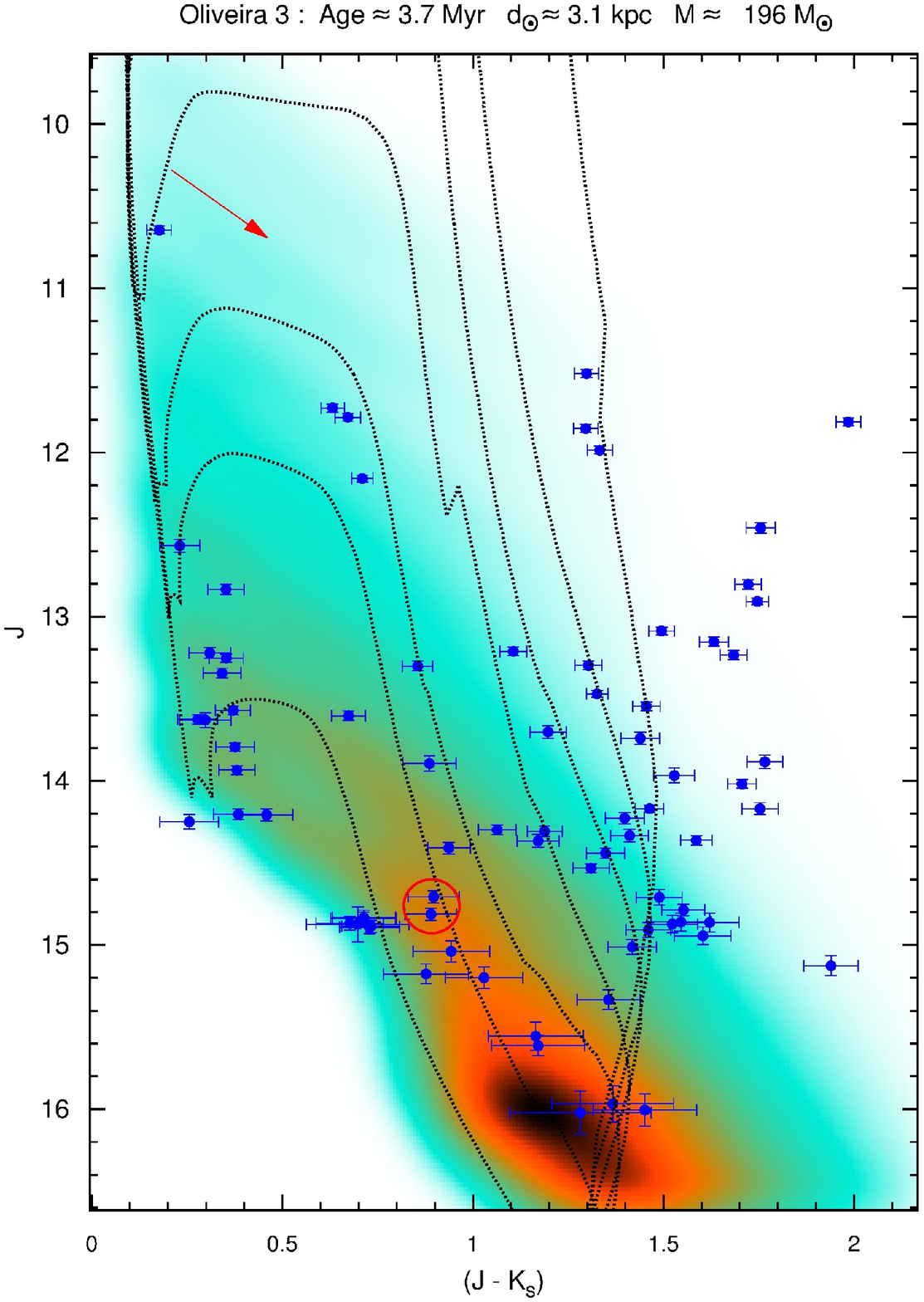}}
\vspace*{-0.2cm}\caption[]{2MASS field-star decontaminated $J\times(J-K_s)$ CMDs of Oliveira\,1 (with isochrones ranging from 0.01 Myr to 2.00$\,$Myr) and Oliveira\,3 (isochrones from 0.01 to 4.00$\,$Myr). The differential reddening vector is also presented. Blue dots are the observed members, dashed lines are PARSEC isochrones shifted according to the derived parameters, and Hess diagram is the simulated model. Dots with red circles correspond to  YSO stars in common, using two catalogues from the literature (see Sect.~\ref{sec:3.3}).}
\label{f4}
\end{figure*}

\begin{figure*}
\resizebox{0.41\textwidth}{!}{\includegraphics{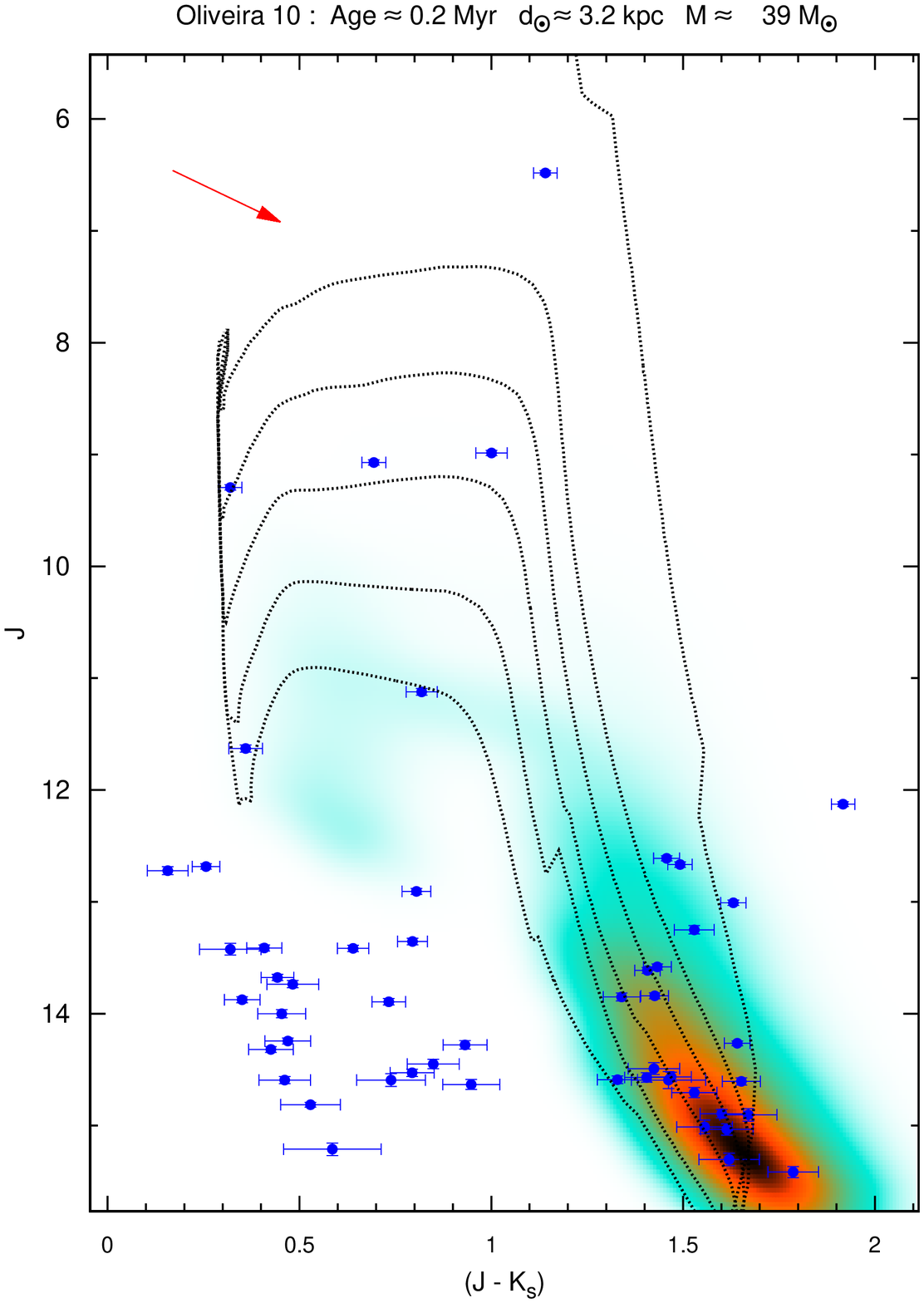}}
\hspace{0.5cm}\resizebox{0.41\textwidth}{!}{\includegraphics{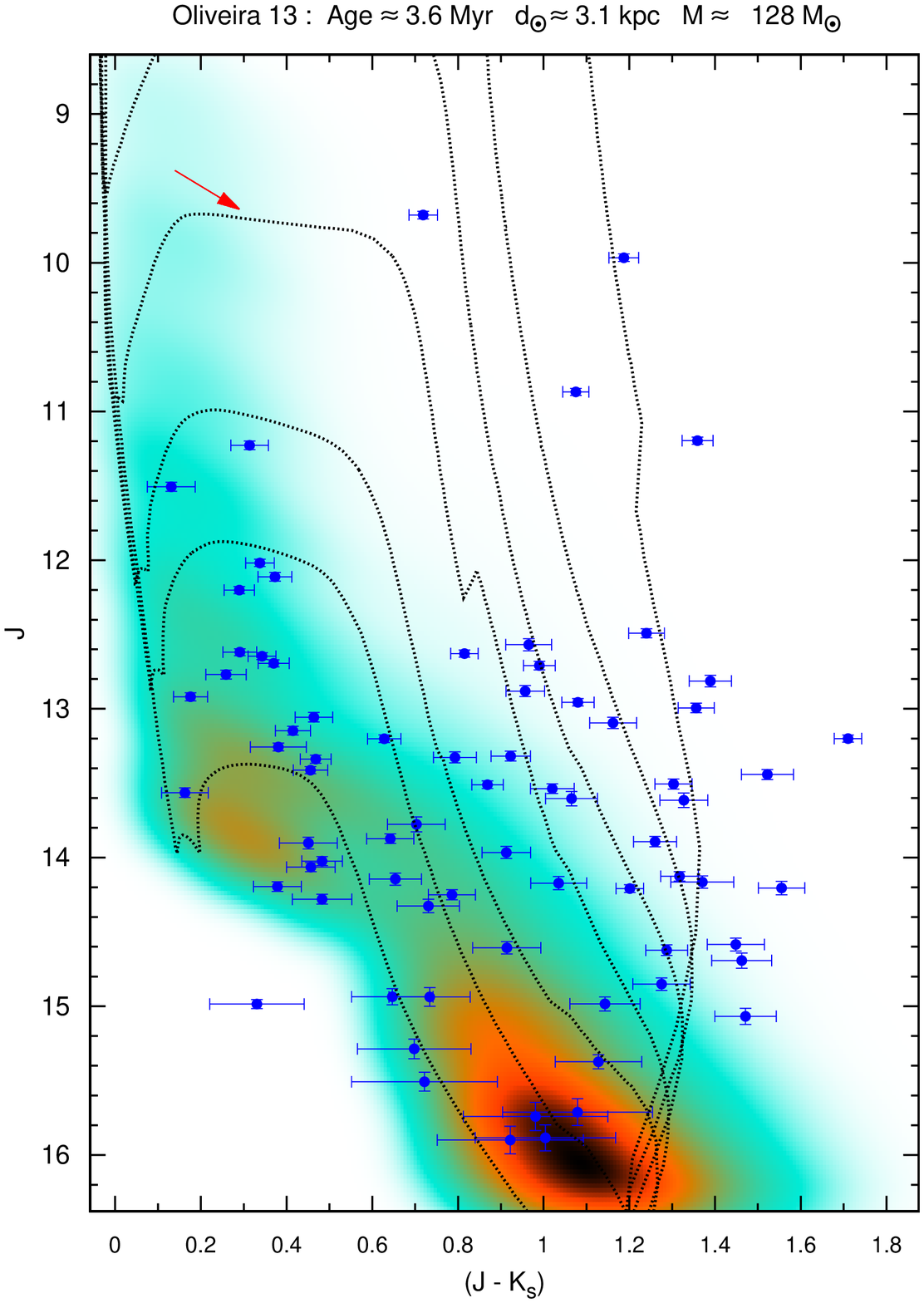}}
\vspace*{-0.2cm}\caption[]{2MASS decontaminated $J\times(J-K_s)$ CMDs of Oliveira\,10 (with isochrones ranging from 0.01 Myr to 0.20$\,$Myr) and Oliveira\,13 (isochrones from 0.01 to 4.00$\,$Myr). Details as in Fig.~\ref{f4}.}
\label{f5}
\end{figure*}

\begin{figure*}
    \vspace{0.3cm}
    \resizebox{0.253\textwidth}{!}{\includegraphics{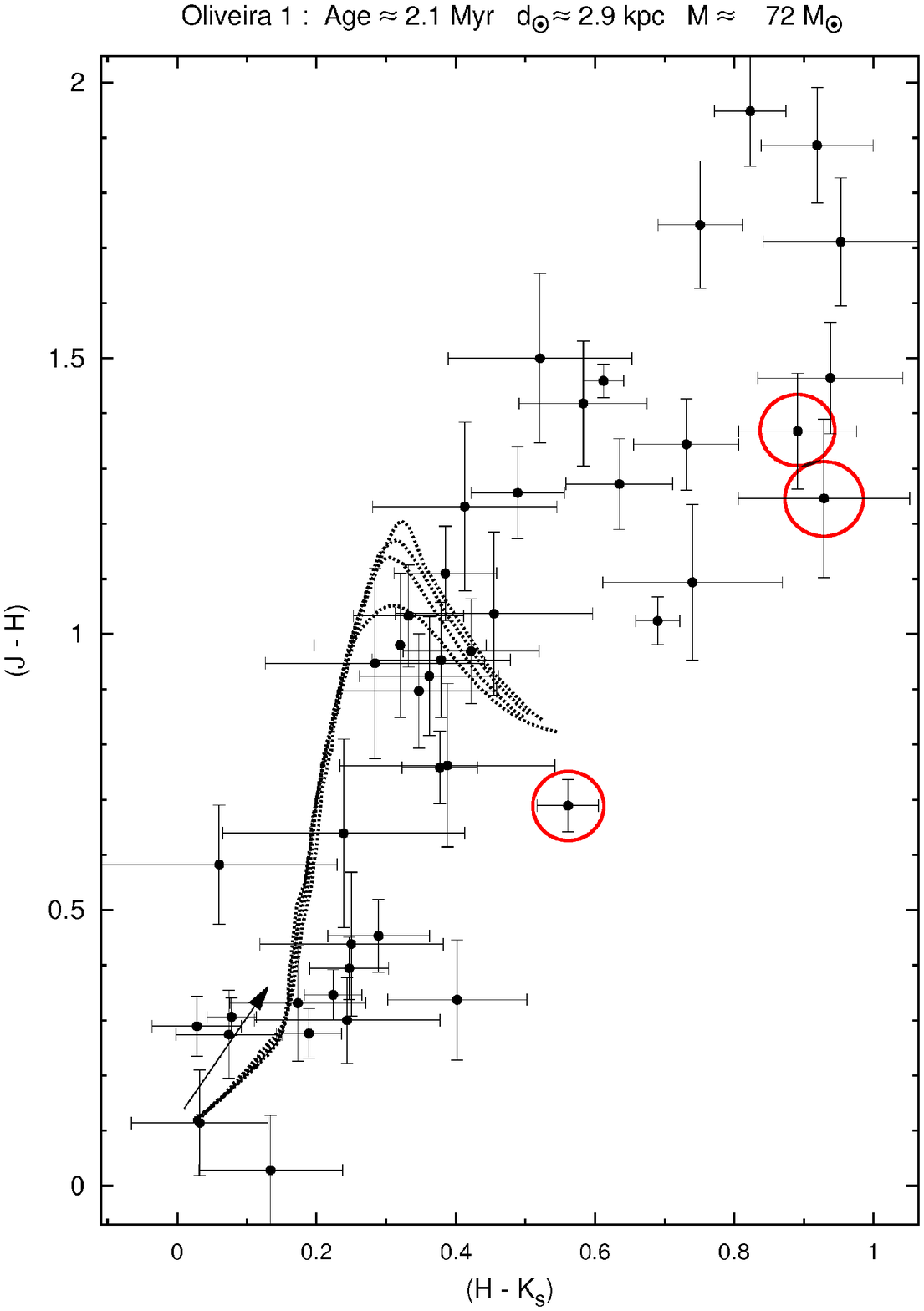}}
    \hspace{-0.18cm}\resizebox{0.253\textwidth}{!}{\includegraphics{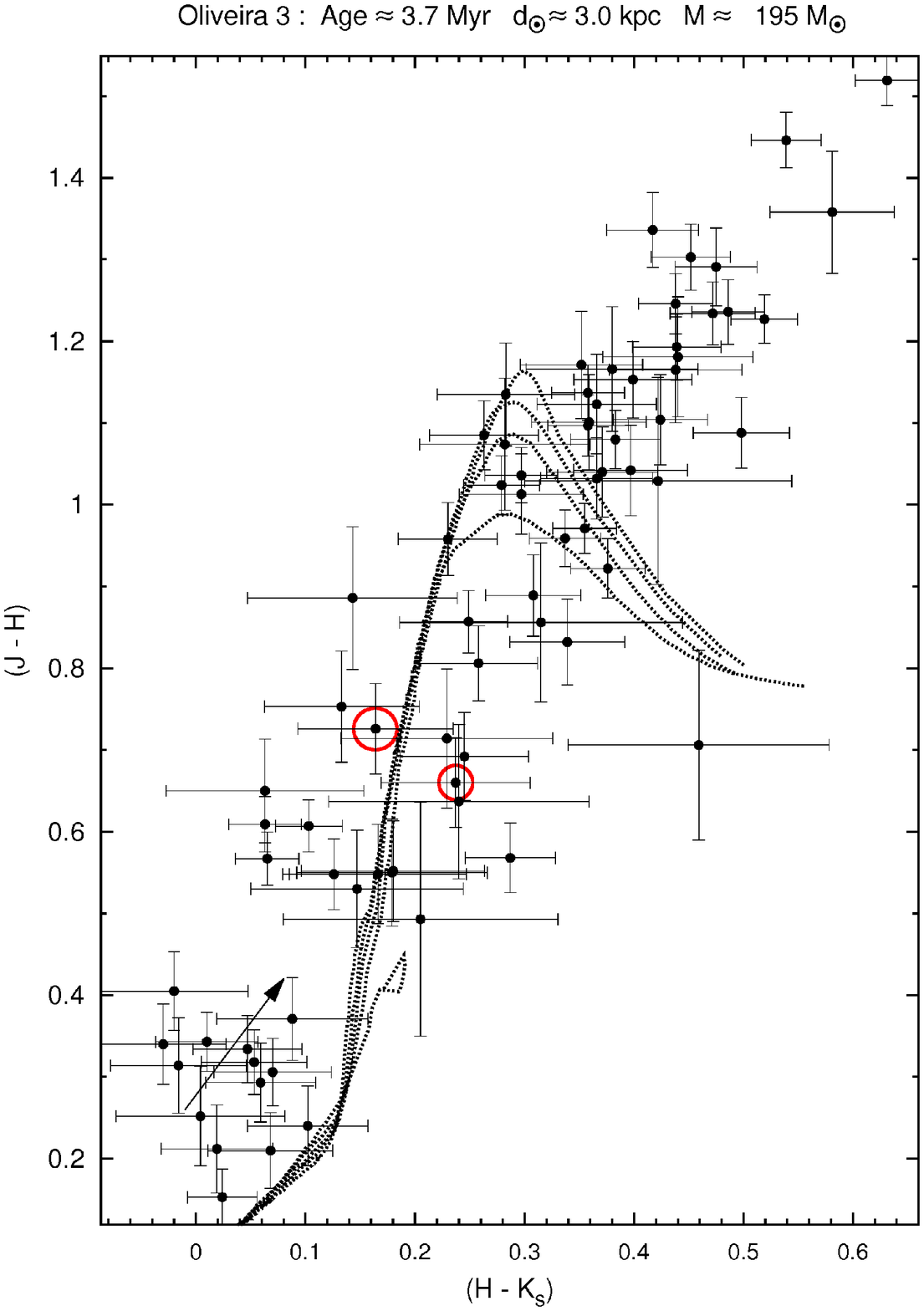}}
    \hspace{-0.18cm}\resizebox{0.253\textwidth}{!}{\includegraphics{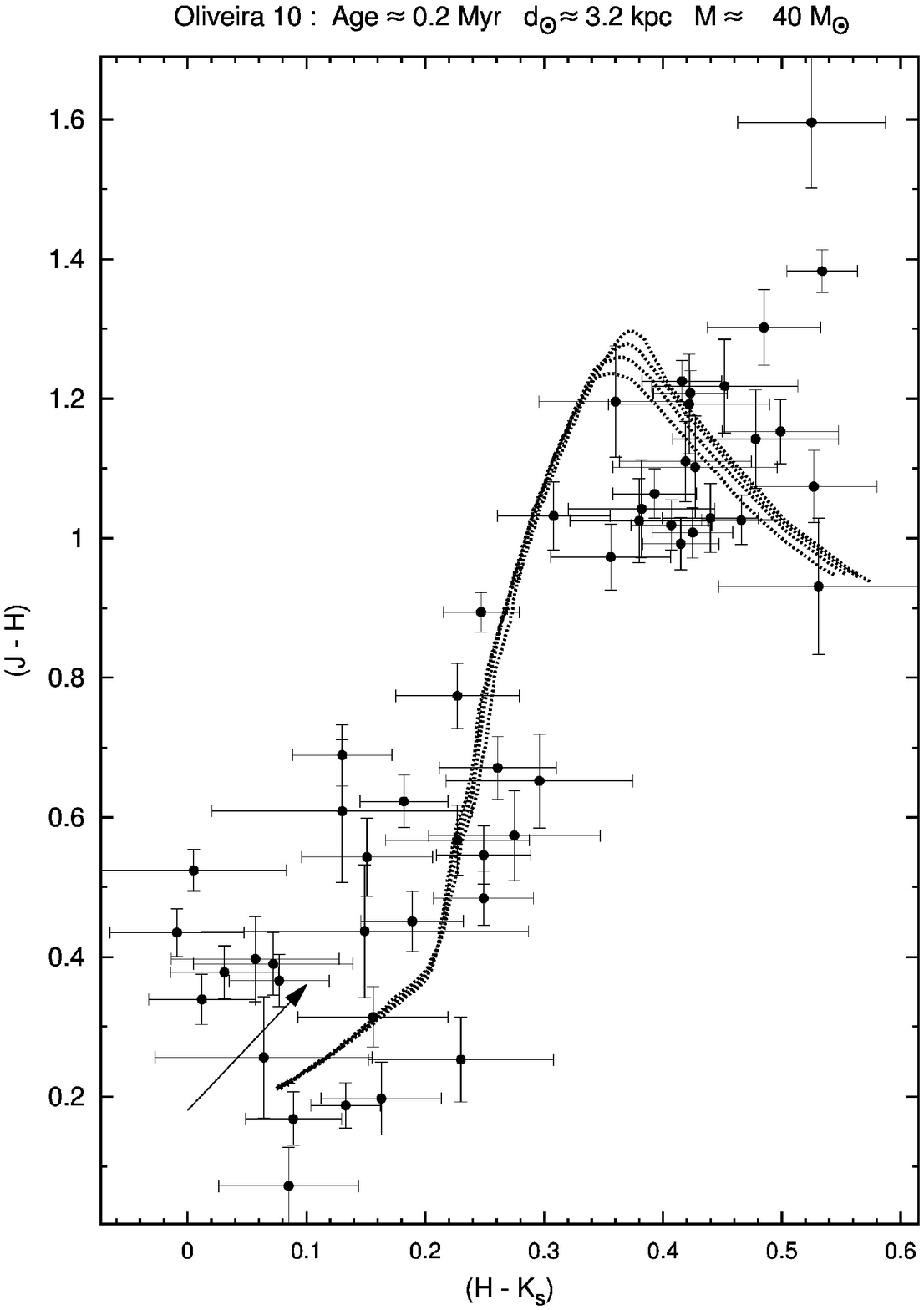}}
    \hspace{-0.18cm}\resizebox{0.253\textwidth}{!}{\includegraphics{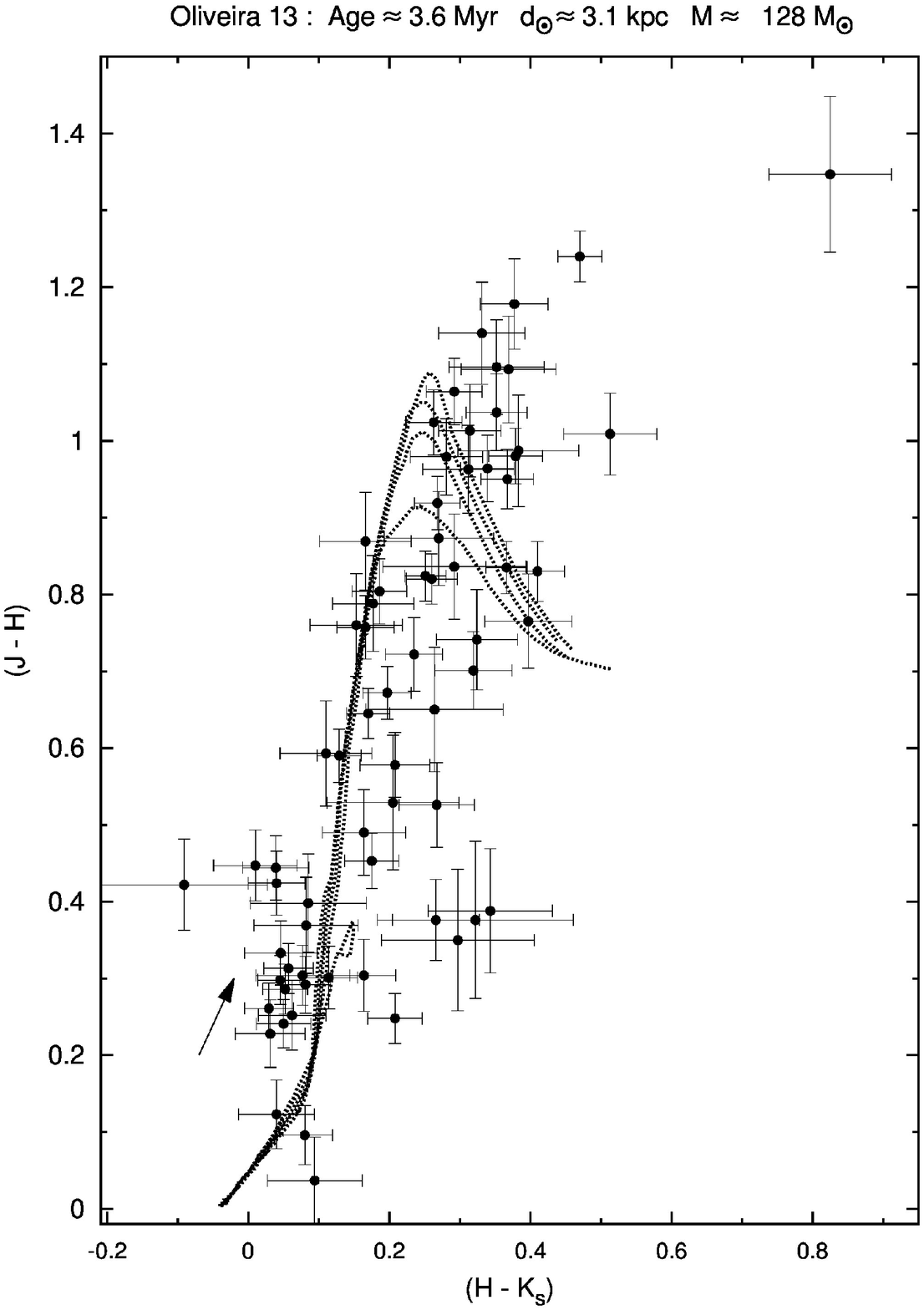}}
    \vspace*{-0.2cm}\caption[]{2MASS $(J-H)\times(H-K_s)$ colour-colour diagrams of Oliveira\,1, Oliveira\,3, Oliveira\,10 and Oliveira\,13. The dots (with colour uncertainties limited to 0.1 mag), reddening vector and PARSEC isochrones as in Fig.~\ref{f4}. Dots with red circles correspond to  YSO stars in common, using two catalogues from the literature (see Sect.~\ref{sec:3.3}).}
    \label{f6}
\end{figure*}

\begin{figure}
\begin{center}
\textcolor{white}{.......}\includegraphics[width=5.4cm]{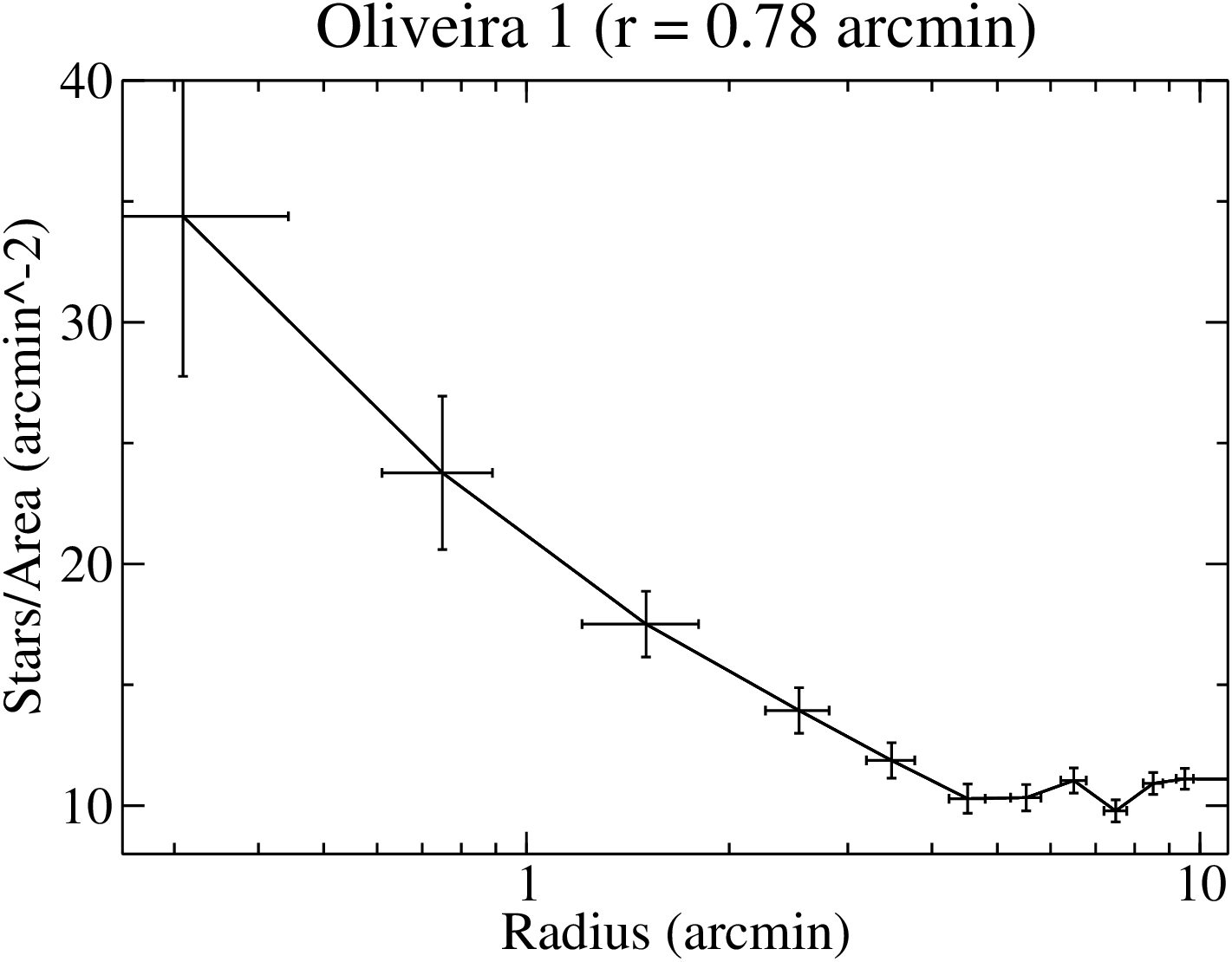}\hspace{0.18cm}
\includegraphics[width=5.4cm]{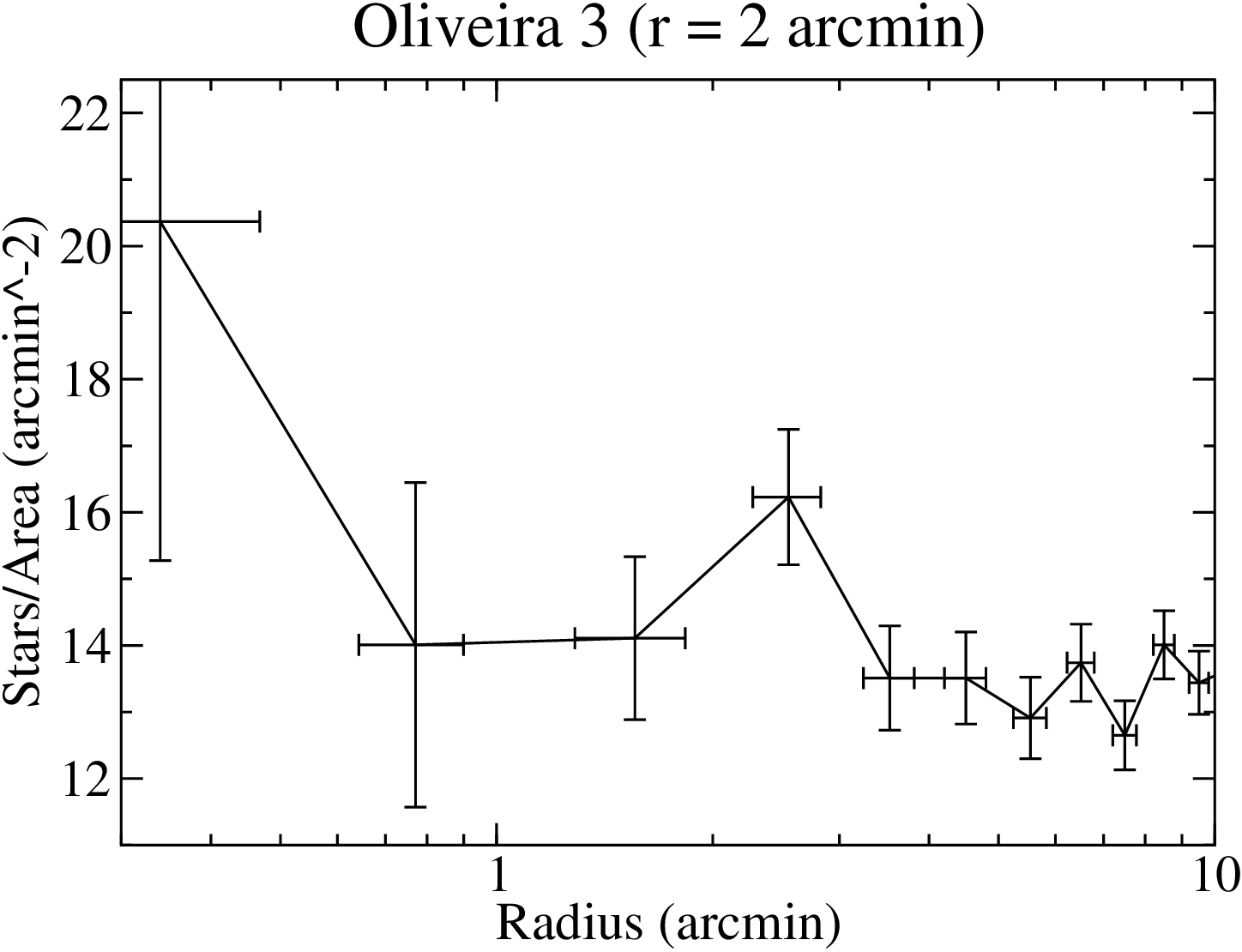}\hspace{0.18cm}
\includegraphics[width=5.4cm]{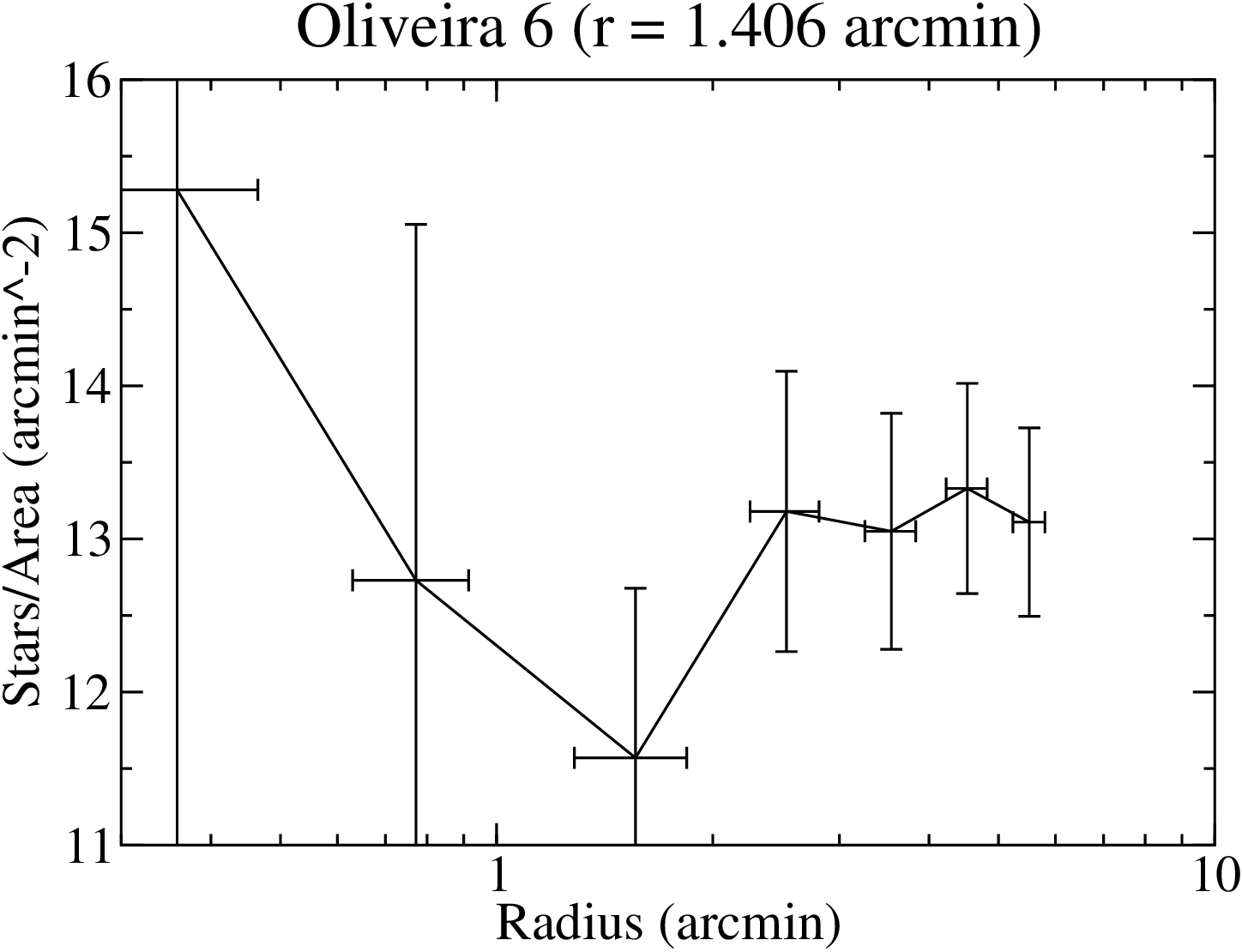}\hspace{0.18cm}\hfill
\vspace{0.3cm}
\textcolor{white}{.......}\includegraphics[width=5.4cm]{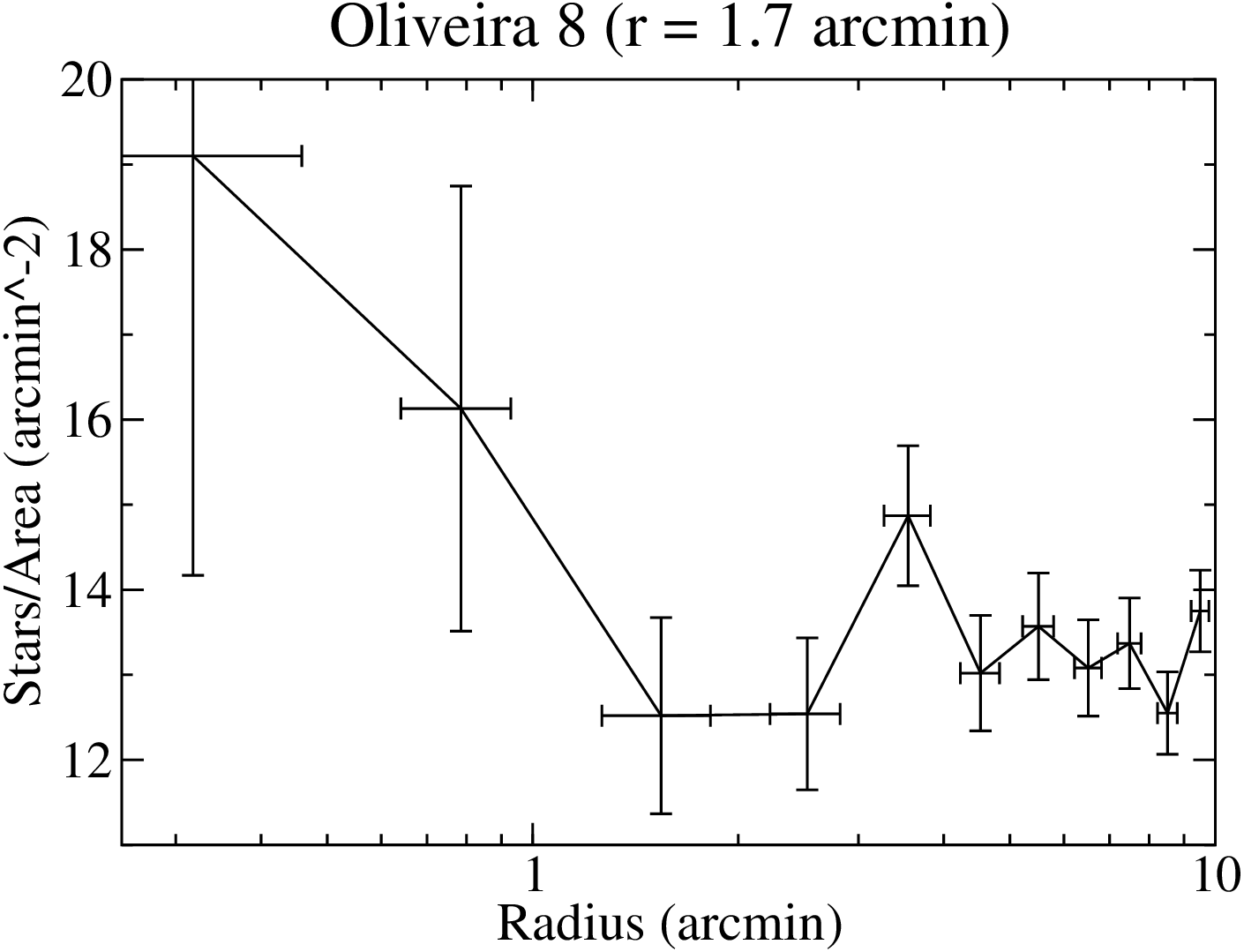}\hspace{0.18cm}
\includegraphics[width=5.4cm]{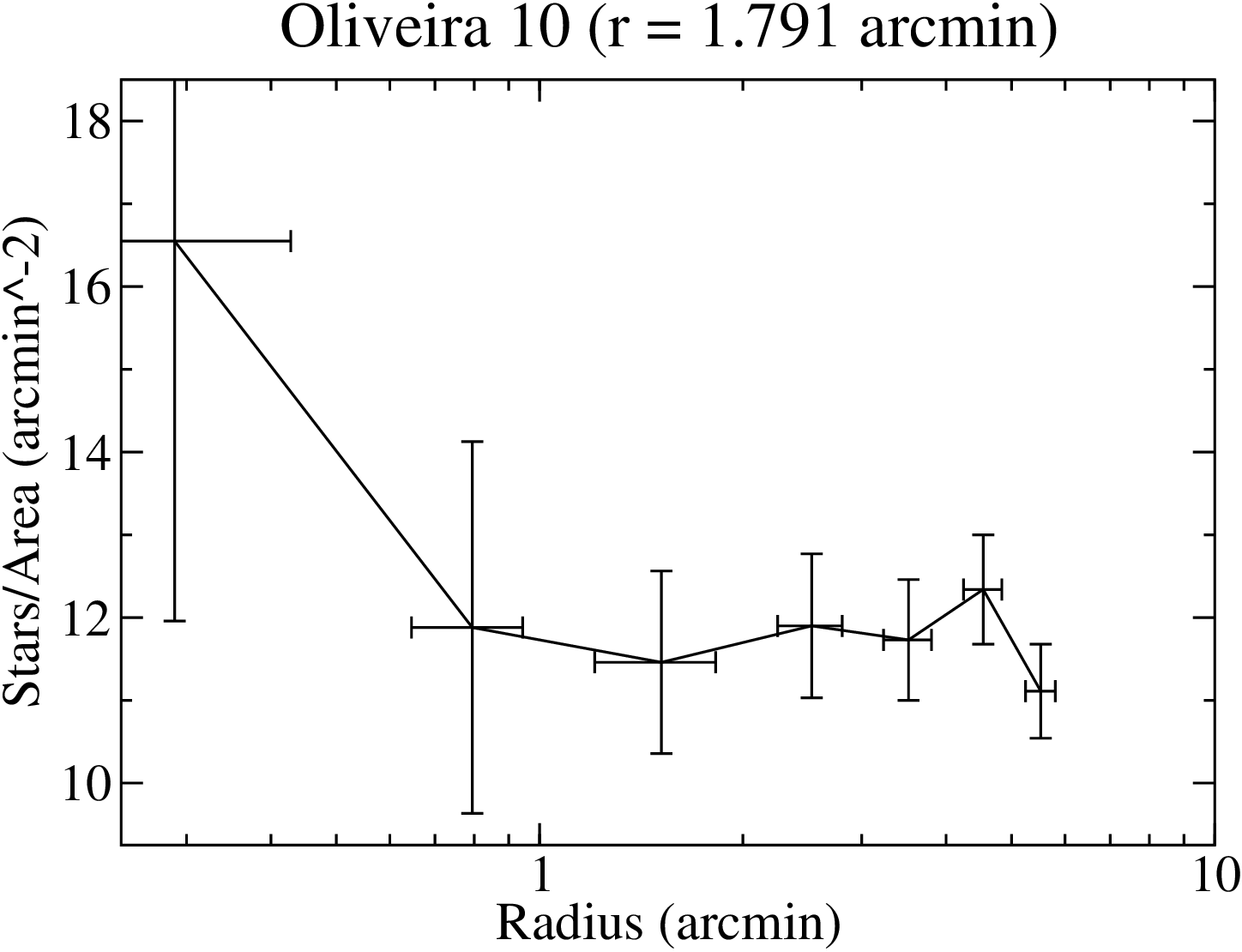}\hspace{0.18cm}
\includegraphics[width=5.4cm]{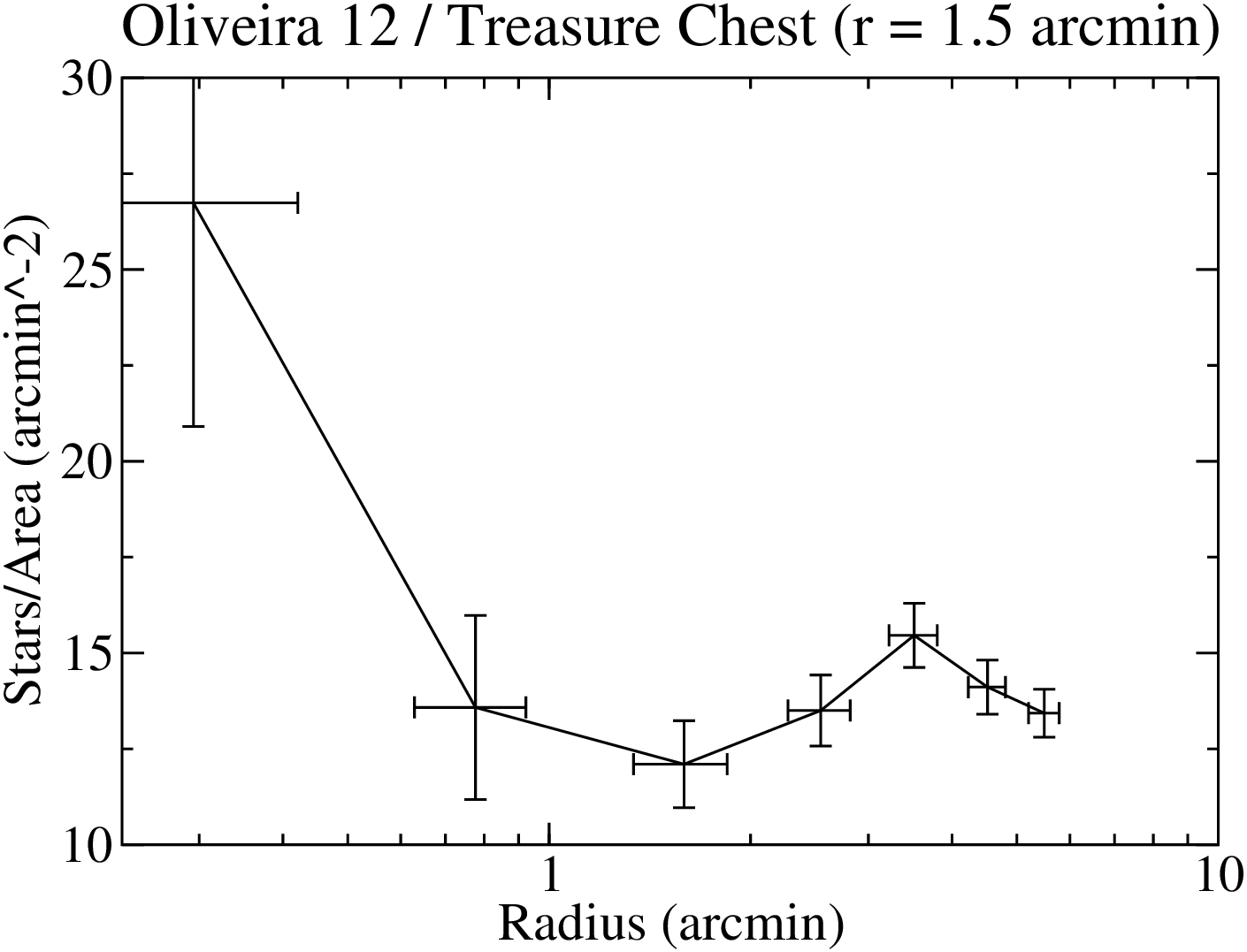}\hspace{0.18cm}\hfill
\vspace{0.3cm}
\textcolor{white}{.......}\includegraphics[width=5.4cm]{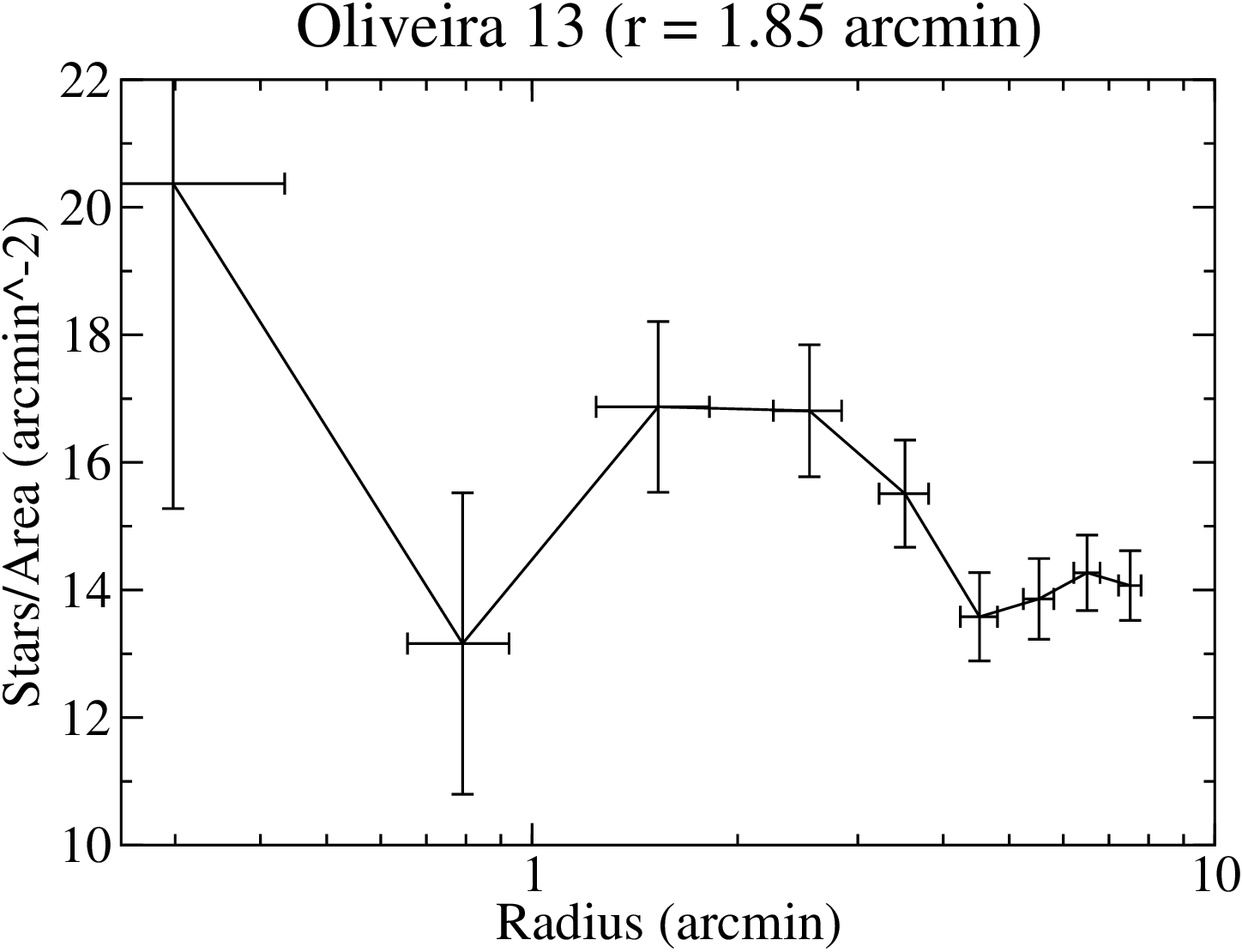}\hspace{0.18cm}
\includegraphics[width=5.4cm]{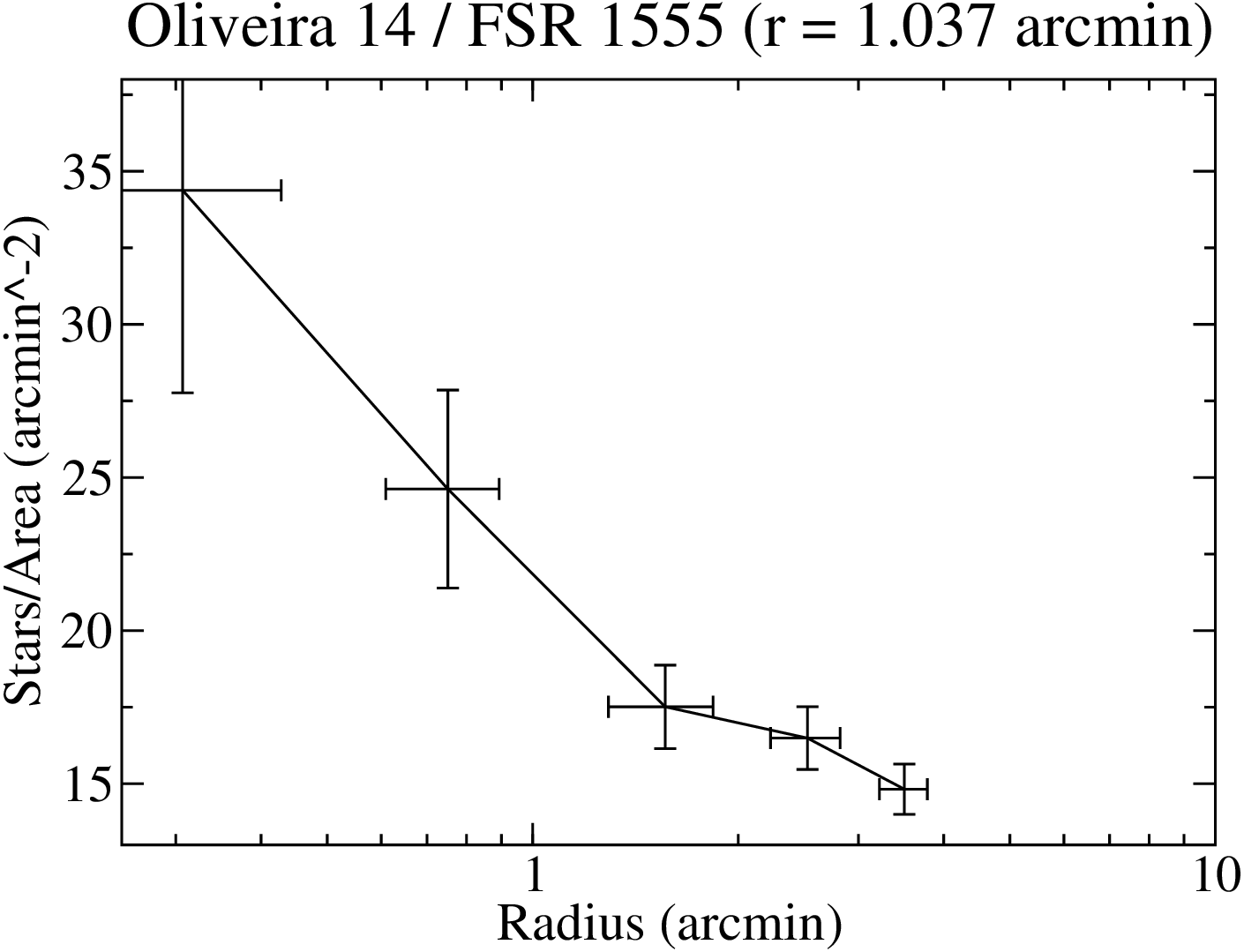}\hspace{0.18cm}
\includegraphics[width=5.4cm]{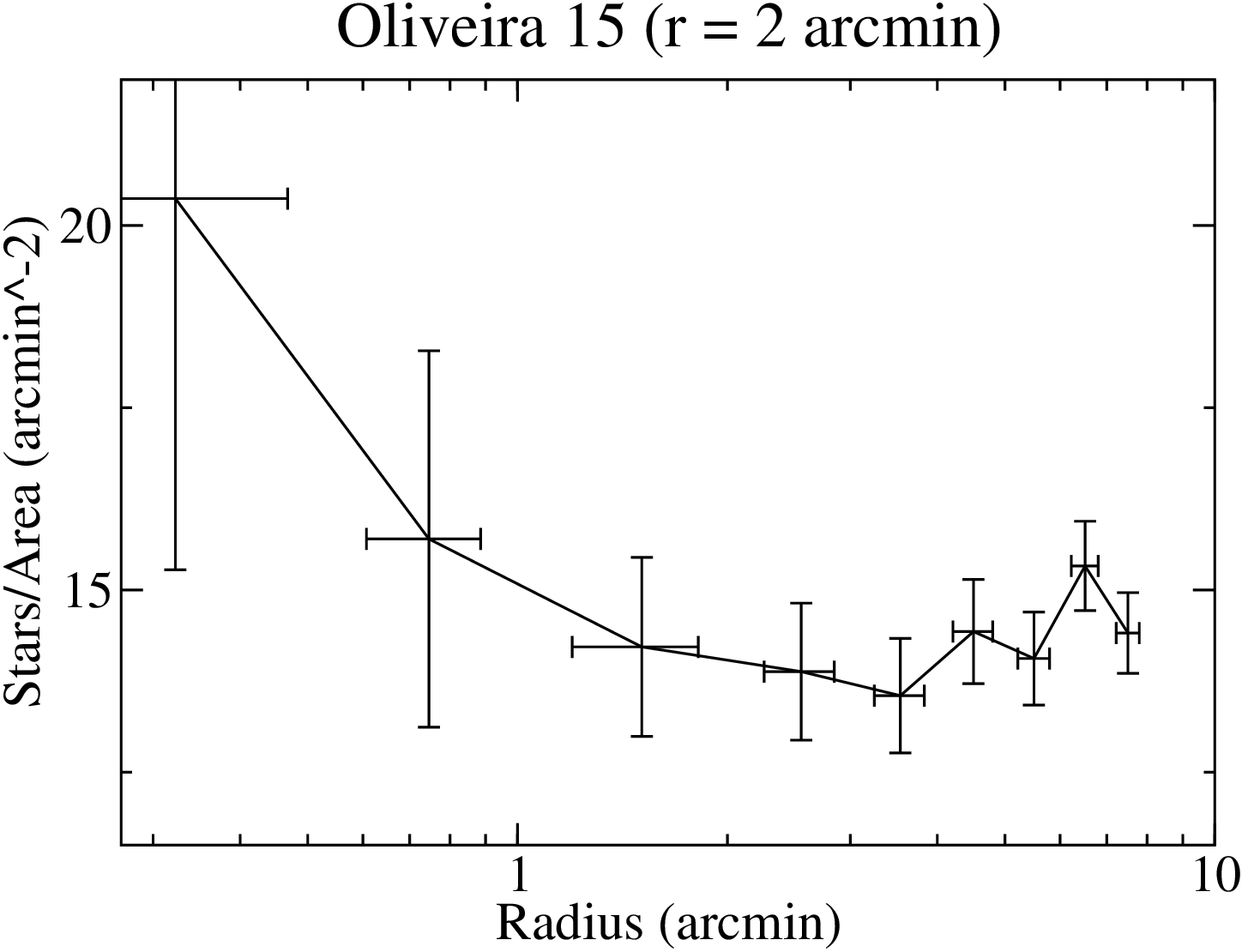}\hspace{0.18cm}\hfill
\vspace*{-0.1cm}\caption{RDPs of the clusters with a central peak, even if small (category $i$). The radius axis is in logarithmic scale.}
\label{f7}
\vspace{0.4cm}
\textcolor{white}{.......}\includegraphics[width=5.4cm]{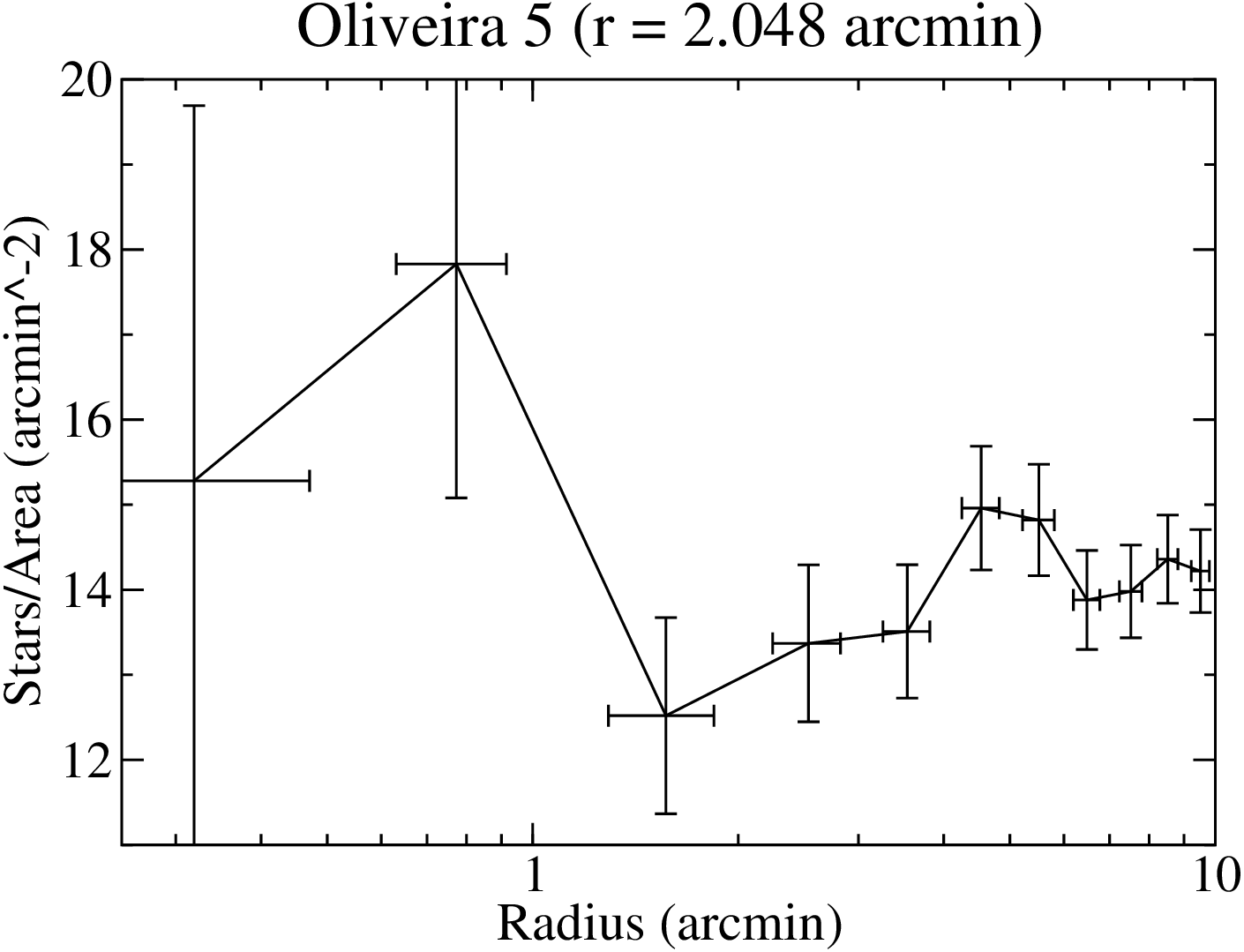}\hspace{0.18cm}
\includegraphics[width=5.4cm]{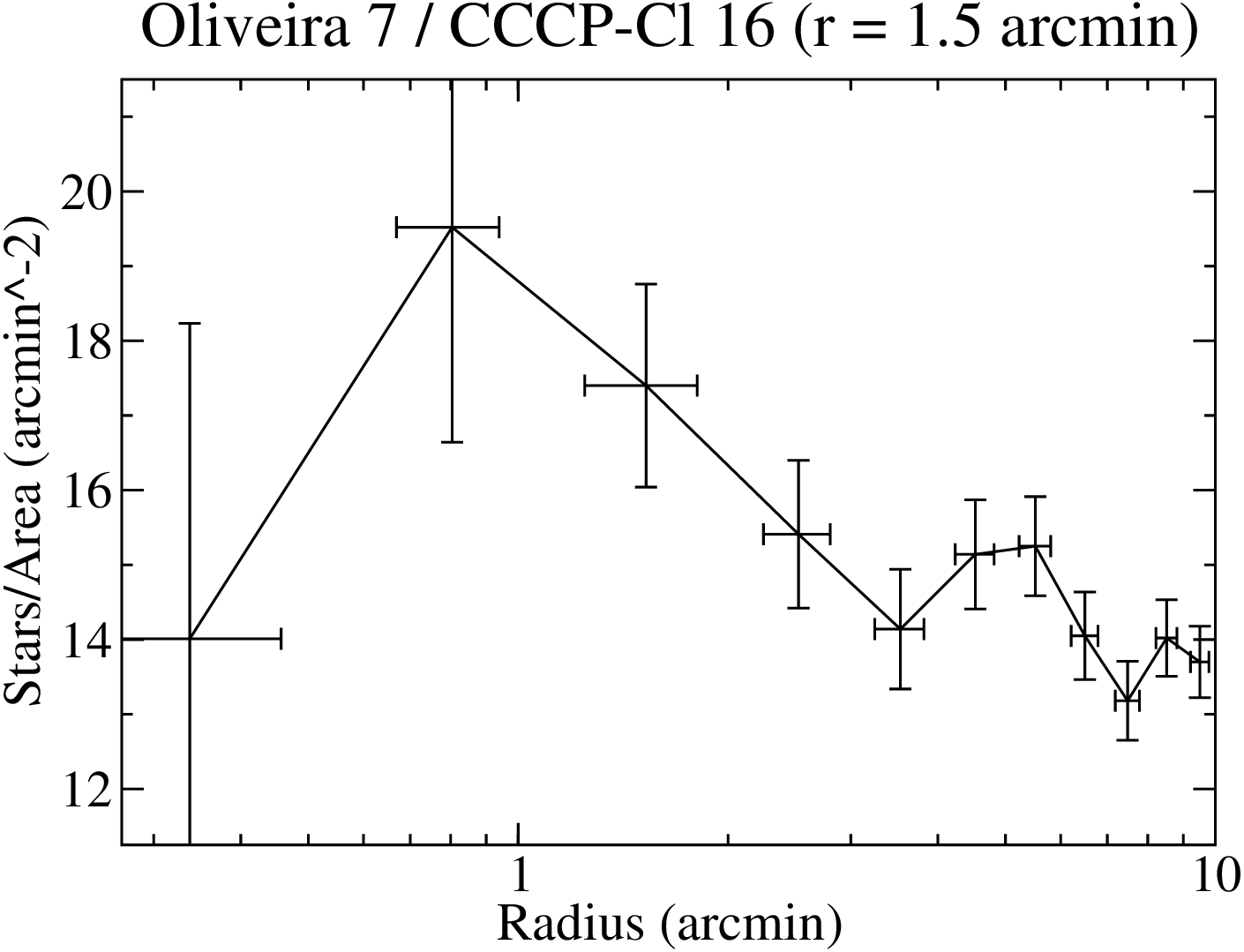}\hspace{0.18cm}
\includegraphics[width=5.4cm]{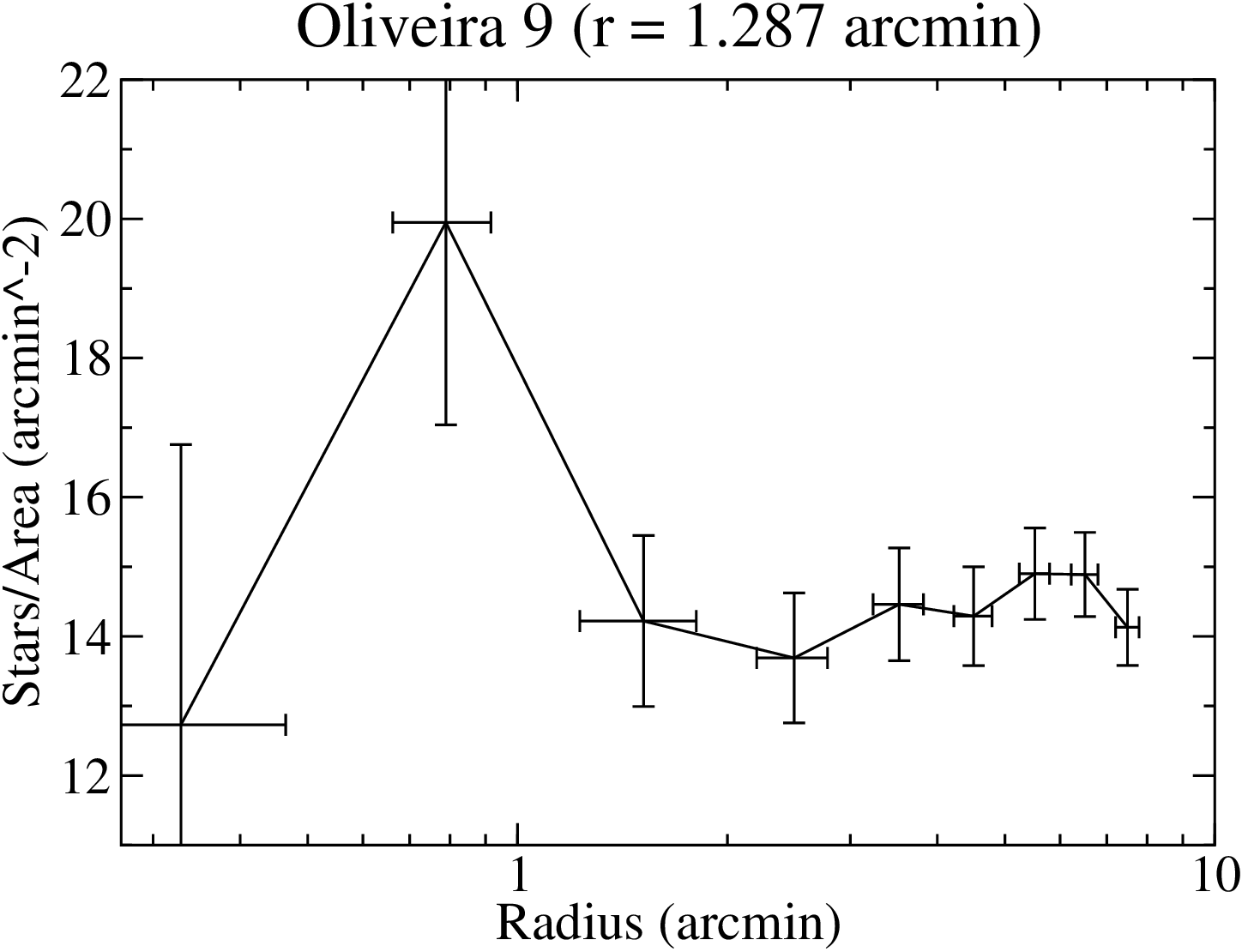}\hspace{0.18cm}\hfill
\vspace{0.3cm}
\textcolor{white}{......}\includegraphics[width=5.4cm]{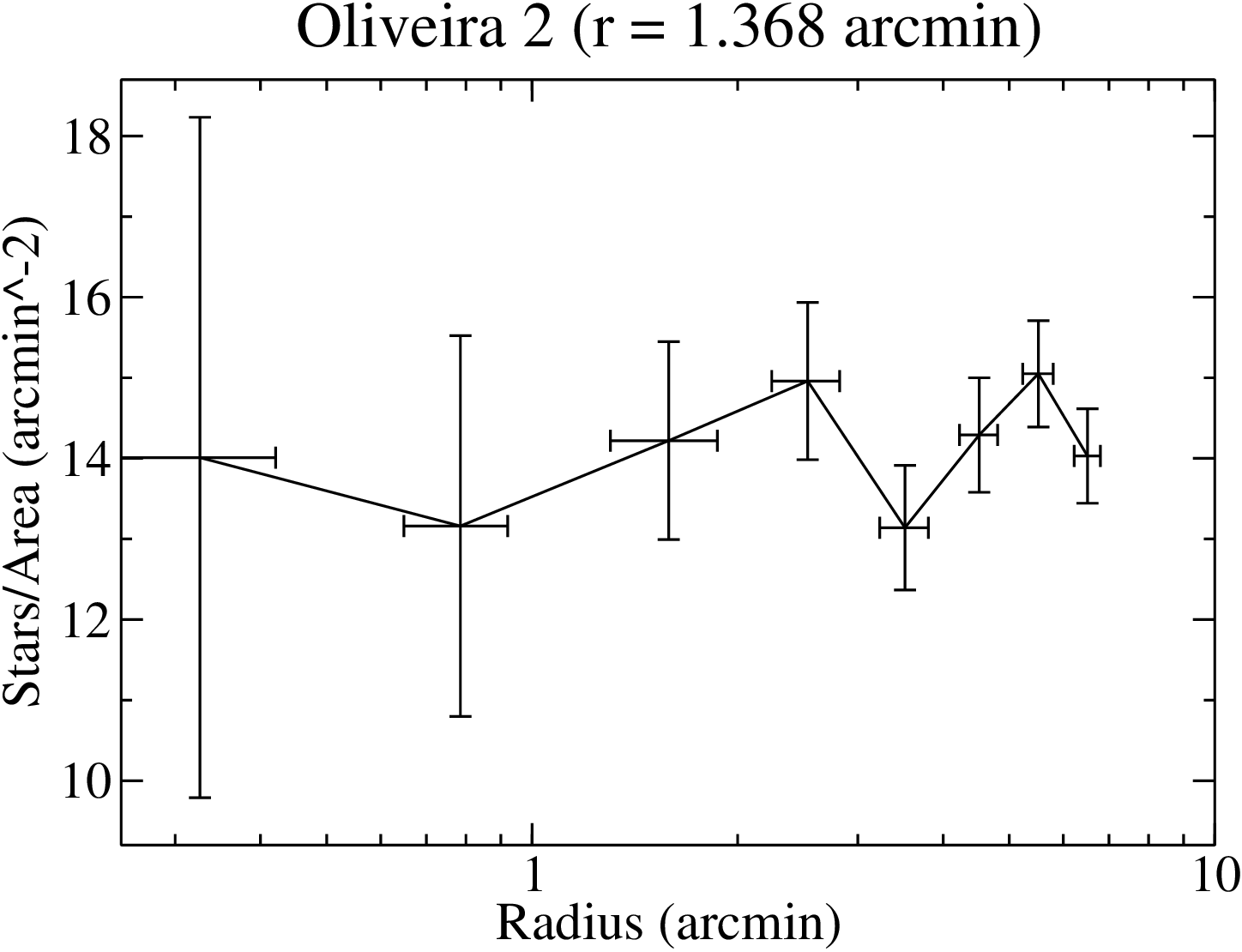}\hspace{0.18cm}
\includegraphics[width=5.4cm]{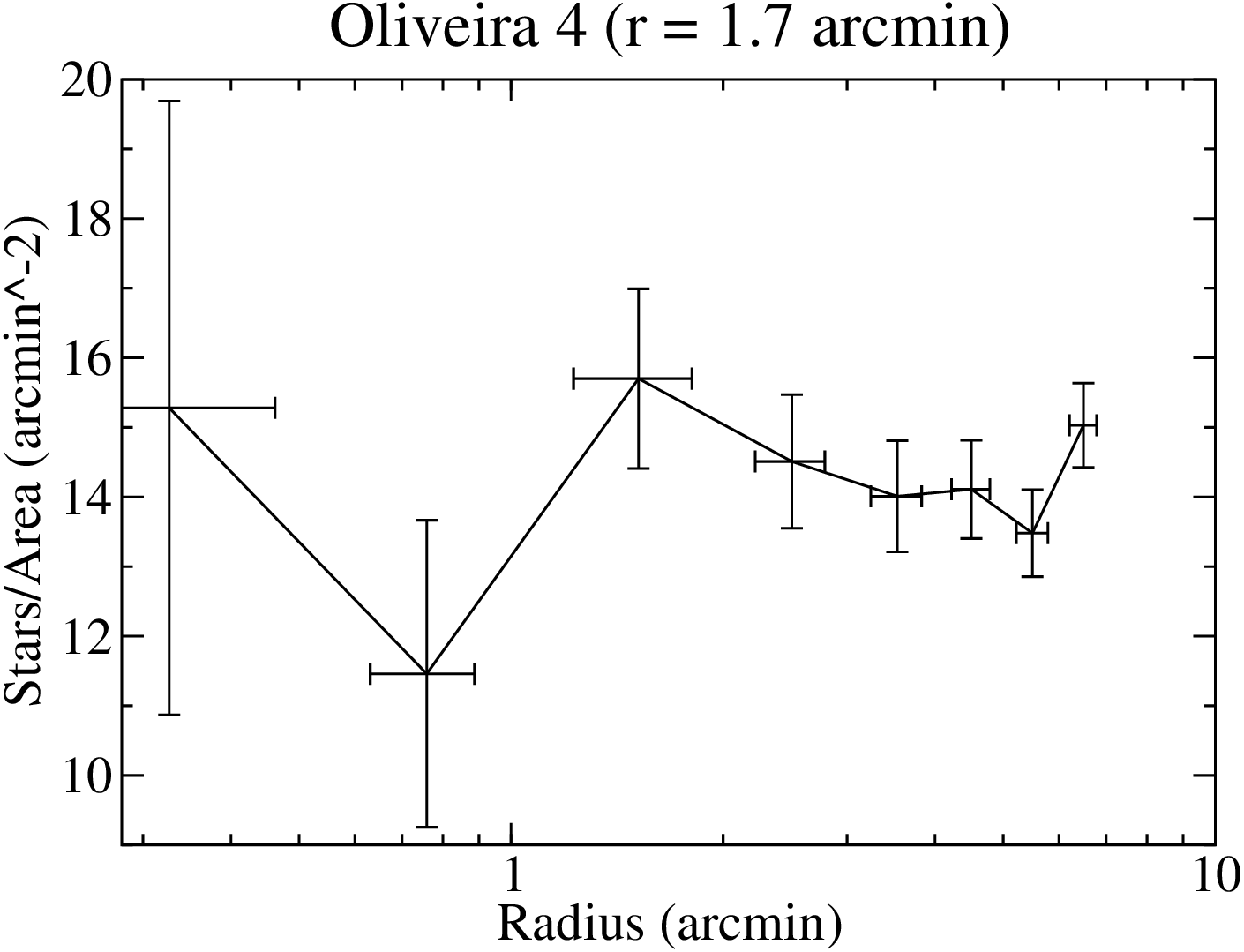}\hspace{0.18cm}
\includegraphics[width=5.4cm]{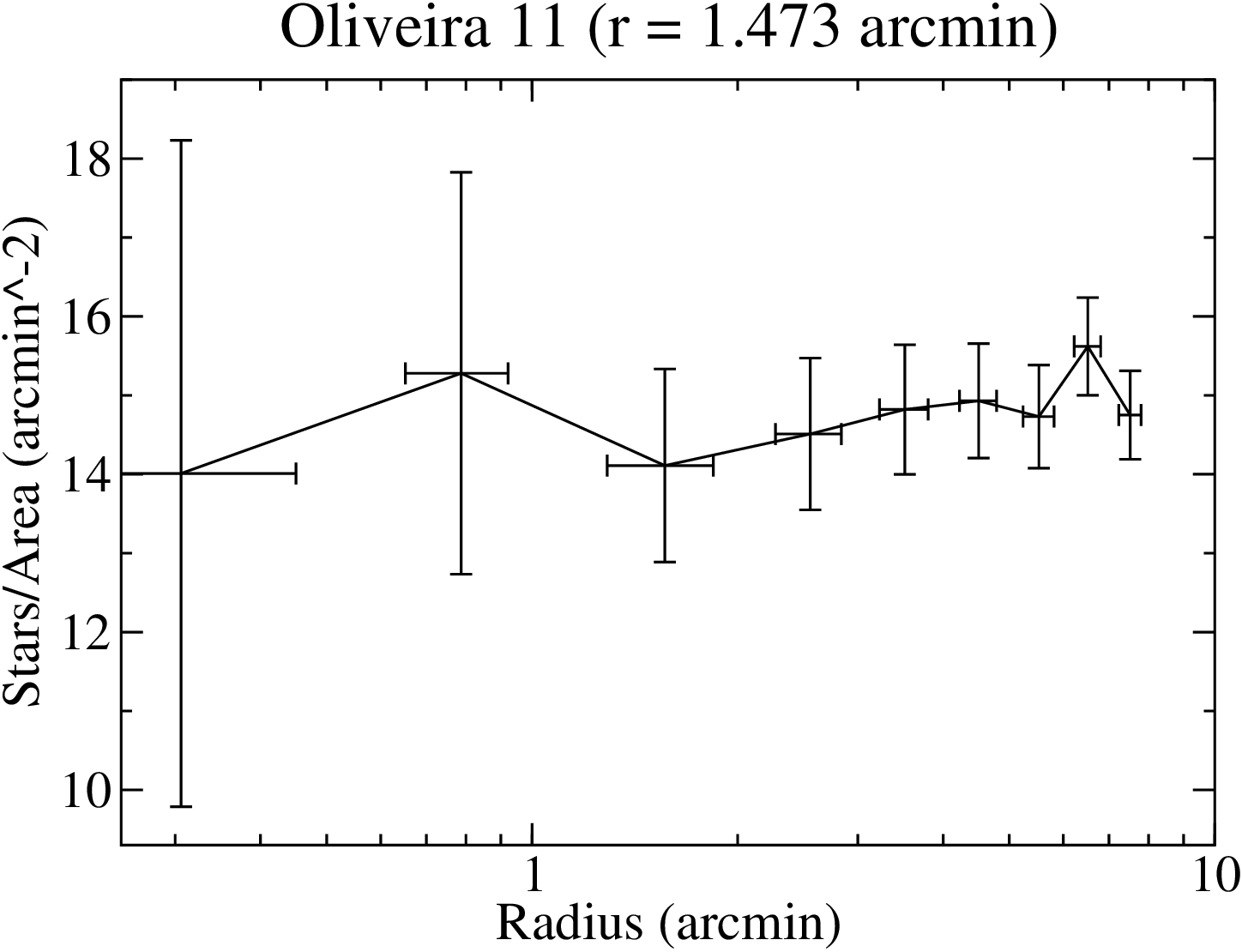}\hspace{0.18cm}\hfill
\vspace*{-0.1cm}\caption[]{Upper panel: RDPs with a non-central peak (category $ii$). Lower panel: Irregular profiles (category $iii$).}
\label{f8}
\end{center}
\end{figure}

Figs.~\ref{f4} and \ref{f5} show colour-magnitude diagrams for some of the new embedded clusters in the present sample. Their colour-colour diagrams are presented in Fig.~\ref{f6}. Additional ones are given in the electronic figures (Figs. 15 to 20). The parameters of the minimum solution are shown in Table~\ref{tab2}, followed by their respective uncertainties, not symmetrical relatively to the central value. All radial density profiles are given in Figs.~\ref{f7} and \ref{f8}.

Figs.~\ref{f4} and \ref{f5} show how the most populated areas of PMS stars in the CMDs have counterpart in the landscape of the computed Hess model. The intrinsic CMDs provide good isochrone fittings and parameters consistent with initial expectations, i.e., ages not above 4.5 Myr (coherent with the embedded phase for a cluster, \citealp{Lada03}) and distances ranging from 2.5 to 3.4 kpc, very similar to the estimated distance of the Carina complex (2.3 to 3.1 kpc).

Analysing the evolutionary stage of the stars in the CMDs, most stars are in pre-main sequence (PMS) stage, very affected by differential reddening. The clusters have a poorly-populated vertical main sequence (MS), which confirms that they are very young. Some of the clusters (e.g. Oliveira\,3, Oliveira\,4 and Oliveira\,15) have a few more stars along the MS. A large number of faint and reddened PMS stars are located mainly in two density wells, for $M_J \gtrsim 13$ mag, with one of them more populated than the other. The redder stars horizontally beyond the isochrones, with higher values of $(J-K_s)$ colour, are expected and correspond to medium-mass stars involved by dust envelopes.

In some CMDs (Oliveira\,2, Oliveira\,6 and Oliveira\,10), we see that some blue stars are found outside the simulated Hess diagram and are not involved by the isochrones. For instance, in the left panel of Fig.~\ref{f5}, the CMD has them around $M_J \approx 14$ mag and $(J-Ks) \approx 0.5$ mag. These blue stars are essentially stars that remain after the decontamination, and might belong to the clusters. We checked that the redder stars of the Oliveira\,10 CMD are in the dust shell that involves the cluster (see Fig.~\ref{f3}), while bluer stars are located in the central region, less affected by dust.

In general, the differential reddening values are high, ensuring that these clusters are still immersed in their parental cloud. In this sense, the 2MASS photometry was extremely useful, since the infrared radiation is much less affected.

The colour-colour diagrams also yielded good results for the embedded clusters, showing that the isochrone fittings were successful and that the PMS stars are highly affected by differential reddening, just as we concluded from CMD analyses. These diagrams present stars in the T Tauri locus \citep{Lima14} and reddened T Tauri stars (clusters in Fig.~\ref{f6}, and also some clusters in the electronic figures, such as Oliveira\,2, CCCP-Cl\,16 and Oliveira\,15).

The catalogued clusters allow us to compare some parameters with the literature. \citet{Feigelson11} provide, for CCCP-Cl\,16, a visual absorption $A_V \approx 4$ mag, which is related with the colour excess $E(J-K_s)$ by $A_V =R_V\times E(B-V)$, with $R_V = 3.1$, and $E(J-K_s)/E(B-V) = 0.52$ \citep{Cardelli89}. Our foreground absorption is very low, but we obtain a high differential reddening $\sigma_{DR} = 4\,$mag. However, those authors claim that their absorptions values are estimated with low precision, from gas column densities, through Chandra.

\clearpage

\begin{table*}
\vspace{0.1cm}
{\footnotesize
\caption{Derived fundamental parameters for embedded clusters in the present study, from 2MASS decontamined photometry.}
\renewcommand{\tabcolsep}{2.1mm}
\renewcommand{\arraystretch}{1.3}
\begin{tabular}{lrrrrrrr}
\hline
\hline
Cluster&Age&$d_{Sun}$&$M_{tot}$&$M_{ext}$&$E(J-Ks)$&$\overline{A_V}(DR)$&$\sigma_{DR}$\\
&(Myr)&(kpc)&(M$_\odot$)&(M$_\odot$)&(mag)&(mag)&(mag)\\
($1$)&($2$)&($3$)&($4$)&($5$)&($6$)&($7$)&($8$) \\
\hline

Oliveira\,1 &$2.1\;^{+0.6}_{-0.6}$&$2.9\;^{+0.0}_{-0.3}$&$71.9\;^{+20.9}_{-18.4}$&91\,(44$\:$stars)&$0.40\;^{+0.00}_{-0.11}$&$1.95\;^{+0.03}_{-0.56}$&$2.0\;^{+0.0}_{-0.6}$\\[1.3ex]
Oliveira\,2 &$0.2\;^{+0.1}_{-0.0}$&$2.5\;^{+0.3}_{-0.4}$&$28.5\;^{+8.2}_{-8.2}$&57\,(82$\:$stars)&$0.33\;^{+0.09}_{-0.09}$&$1.27\;^{+0.36}_{-0.36}$&$0.7\;^{+0.2}_{-0.2}$\\[1.3ex]
Oliveira\,3 &$3.7\;^{+0.2}_{-1.1}$&$3.0\;^{+0.1}_{-0.7}$&$195.8\;^{+56.7}_{-56.4}$&220\,(77$\:$stars)&$0.34\;^{+0.10}_{-0.10}$&$1.45\;^{+0.03}_{-0.42}$&$3.0\;^{+0.0}_{-0.9}$\\[1.3ex]
Oliveira\,4 &$4.5\;^{+0.3}_{-1.3}$&$2.9\;^{+0.4}_{-0.3}$&$196.2\;^{+56.6}_{-56.7}$&210\,(58$\:$stars)&$0.38\;^{+0.11}_{-0.02}$&$1.99\;^{+0.01}_{-0.57}$&$2.0\;^{+0.0}_{-0.6}$\\[1.3ex]
Oliveira\,5 &$4.5\;^{+0.3}_{-1.2}$&$3.3\;^{+0.0}_{-0.7}$&$150.8\;^{+43.7}_{-43.6}$&170\,(63$\:$stars)&$0.10\;^{+0.03}_{-0.00}$&$1.73\;^{+0.16}_{-0.50}$&$0.8\;^{+0.2}_{-0.2}$\\[1.3ex]
Oliveira\,6 &$0.3\;^{+0.1}_{-0.0}$&$2.7\;^{+0.5}_{-0.1}$&$14.3\;^{+4.2}_{-2.5}$&31\,(37$\:$stars)&$0.38\;^{+0.10}_{-0.05}$&$0.50\;^{+0.14}_{-0.14}$&$0.6\;^{+0.2}_{-0.2}$\\[1.3ex]
CCCP-Cl\,16 &$2.0\;^{+0.6}_{-0.6}$&$2.8\;^{+0.6}_{-0.7}$&$162.2\;^{+46.9}_{-46.9}$&190\,(69$\:$stars)&$0.00\;^{+0.00}_{-0.00}$&$0.35\;^{+0.09}_{-0.10}$&$4.0\;^{+0.0}_{-1.2}$\\[1.3ex]
Oliveira\,8 &$0.4\;^{+0.1}_{-0.1}$&$2.7\;^{+0.6}_{-0.1}$&$37.9\;^{+11.0}_{-10.9}$&71\,(62$\:$stars)&$0.27\;^{+0.08}_{-0.08}$&$0.40\;^{+0.06}_{-0.12}$&$3.0\;^{+0.0}_{-0.9}$\\[1.3ex]
Oliveira\,9 &$2.6\;^{+0.2}_{-0.7}$&$2.7\;^{+0.1}_{-0.8}$&$172.7\;^{+50.1}_{-50.0}$&200\,(79$\:$stars)&$0.40\;^{+0.12}_{-0.11}$&$1.99\;^{+0.00}_{-0.58}$&$2.5\;^{+0.7}_{-0.7}$\\[1.3ex]
Oliveira\,10 &$0.2\;^{+0.1}_{-0.1}$&$3.2\;^{+0.1}_{-0.5}$&$39.5\;^{+11.4}_{-11.2}$&64\,(55$\:$stars)&$0.54\;^{+0.09}_{-0.08}$&$1.61\;^{+0.22}_{-0.46}$&$0.7\;^{+0.2}_{-0.2}$\\[1.3ex]
Oliveira\,11 &$3.9\;^{+0.5}_{-0.5}$&$3.3\;^{+0.1}_{-1.5}$&$92.4\;^{+26.8}_{-26.8}$&110\,(39$\:$stars)&$0.14\;^{+0.04}_{-0.01}$&$0.06\;^{+0.02}_{-0.02}$&$2.0\;^{+0.0}_{-0.6}$\\[1.3ex]
Treasure Chest &$1.3\;^{+0.4}_{-0.4}$&$2.6\;^{+0.1}_{-0.2}$&$62.6\;^{+18.1}_{-18.1}$&80\,(52$\:$stars)&$0.11\;^{+0.03}_{-0.01}$&$3.67\;^{+0.77}_{-0.39}$&$5.0\;^{+0.0}_{-1.5}$\\[1.3ex]
Oliveira\,13 &$3.6\;^{+0.2}_{-1.0}$&$3.1\;^{+0.1}_{-0.6}$&$127.9\;^{+37.1}_{-36.7}$&150\,(77$\:$stars)&$0.22\;^{+0.06}_{-0.01}$&$0.92\;^{+0.05}_{-0.27}$&$1.0\;^{+0.0}_{-0.3}$\\[1.3ex]
Oliveira\,15 &$3.5\;^{+0.3}_{-0.9}$&$3.4\;^{+0.0}_{-0.6}$&$186.9\;^{+53.9}_{-54.0}$&210\,(71$\:$stars)&$0.20\;^{+0.06}_{-0.00}$&$0.98\;^{+0.01}_{-0.28}$&$1.0\;^{+0.0}_{-0.3}$\\

\hline
\end{tabular}
\begin{list}{Table Notes.}
\item Col. 2: Cluster age. Col. 3: Distance from the Sun. Col. 4: Cluster total mass. Col. 5: Mass extrapolation for the number of cluster members, given between parentheses. Col. 6: Colour excess $E(J-K_s)$. Cols. 7-8: Gaussian's center and dispersion, adjusted for differential reddening. All parameters are detailed in section \ref{sec:2.3}.
\end{list}
\label{tab2}
}
\end{table*}

\clearpage

\onecolumn
%

\twocolumn

\clearpage

For the Treasure Chest, \citet{Smith05} give the following parameters: $0.1$ Myr, $2.3$ kpc and $E(B-V) = 0.65$ mag; in this paper we get: $1.3$ Myr, $2.6$ kpc and $E(B-V) = 0.21$ mag. The excesses are compatible, and the ages and distances are nearly the same. In the paper above, it is also reported that the Treasure Chest stars are highly embedded, presenting high extinction values. In the present paper, we obtain $A_V = 0.65$ mag and a high differential reddening for this cluster, and we also observe several reddened stars, with colours $(J-K_s)$ higher than 2.0 mag.

\subsection{Radial density profiles}
\label{sec:3.2}

The radial density profile provides an overview of the cluster structure and radial distribution of stars. Assuming a Poisson distribution in star counts, the errors in density are given by $\sqrt{N_i}/A_i$, where $N_i$ is the number of stars in each ring and $A_i$ is the ring's area. The uncertainty in radius corresponds to half of the ring thickness. 

According to Figs.~\ref{f7} and \ref{f8}, there are essentially three groups of similar RDPs, as follows: $(i)$ in most cases (objects 1, 3, 6, 8, 10, and from 12 to 15 - Fig.~\ref{f7}) the RDP presents a central density peak, so that they are centrally condensed ECs (in some cases, the King profile may fit well, although they did not reach dynamical relaxation yet); $(ii)$ in other cases (objects 5, 7 and 9 - Fig.~\ref{f8}, upper panel) the peaks are not central, because the higher densities do not occur right at the cluster center; $(iii)$ the remaining cases (objects 2, 4 and 11 - Fig.~\ref{f8}) present several density peaks and low stellar counts, possibly resulting from extrinsic fluctuations due to the dust absorption and to the small number of members.

As one goes to least massive ECs, their profiles become often disturbed. One might question their cluster nature, but their CMDs, colour-colour diagrams and shell structures indicate clusters. We might be dealing with the embedded cluster low-mass limit \citep{Lada03}. In our sample, this lower mass limit corresponds to cluster total stellar masses of 14 to 72 M$_\odot$ (column 4 from Table~\ref{tab2}) and cluster extrapolated masses of 31 to 91 M$_\odot$ (column 5 from Table~\ref{tab2}, with the number of members enclosed in parentheses), the latter including Kroupa\textsc{\char13}s extension.

Irregular profiles (such as for Oliveira 2, 4 and 11 - Fig.~\ref{f8}) certainly reflect a RDP affected not only by stochastic effects, but also by variable absorption and crowding. The counts may be improved in the future by extending the stellar content to masses as low as brown dwarfs. Presently, technical limitations like the 2MASS magnitude cutoff, pixel size and resolution contribute to create these irregular profiles. However, we point out that from the CMD analysis they are ECs, which the present study considers more fundamental than the RDP for classification purposes. This is a concept that we call attention, as future studies will face numerous such low-mass ECs throughout the disk.

\begin{figure}
\centering
\includegraphics[width=6.2cm]{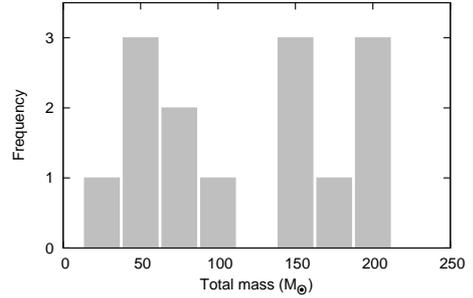}
\vspace*{-0.5cm}\caption[]{Cluster total mass distribution.}
\label{f9}
\end{figure}

\begin{figure}
\vspace{-0.45cm}
\centering
\includegraphics[width=6.2cm]{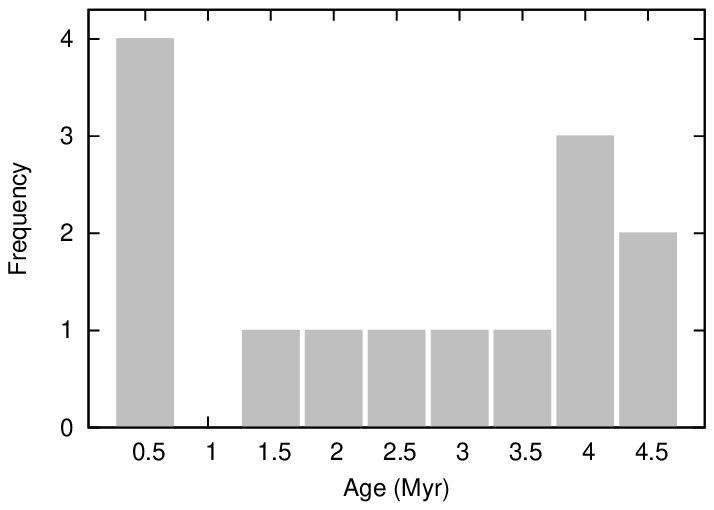}
\vspace*{-0.3cm}\caption[]{Cluster age distribution.}
\label{f10}
\includegraphics[width=6.2cm]{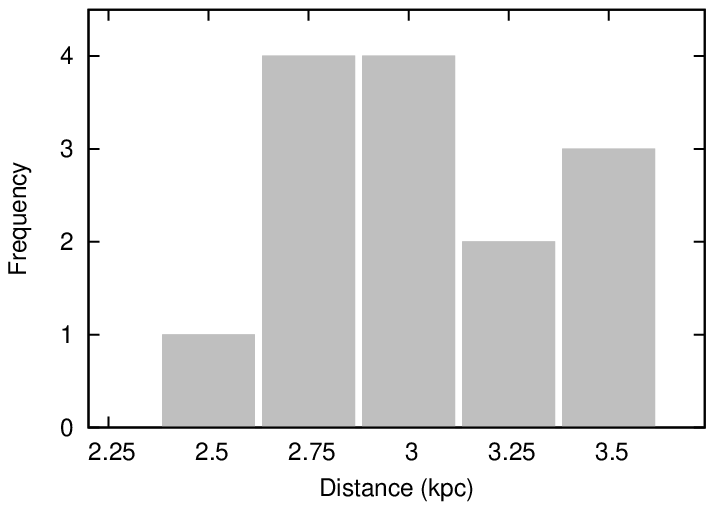}
\vspace*{-0.3cm}\caption[]{Cluster distance distribution.}
\label{f11}
\end{figure}

Figs.~\ref{f9}, \ref{f10} and \ref{f11} show histograms with total masses, ages and distances distributions, respectively, of the 14 ECs. The sample is dominated by low-mass clusters (all under 200\,M$_\odot$). The ages are typical of ECs (Fig.~\ref{f10}). Finally, the distance distribution is compatible with the Carina complex, in the Sagittarius-Carina spiral arm.

\begin{figure}
\centering
\vspace{0.1cm}
\resizebox{0.38\textwidth}{!}{\includegraphics{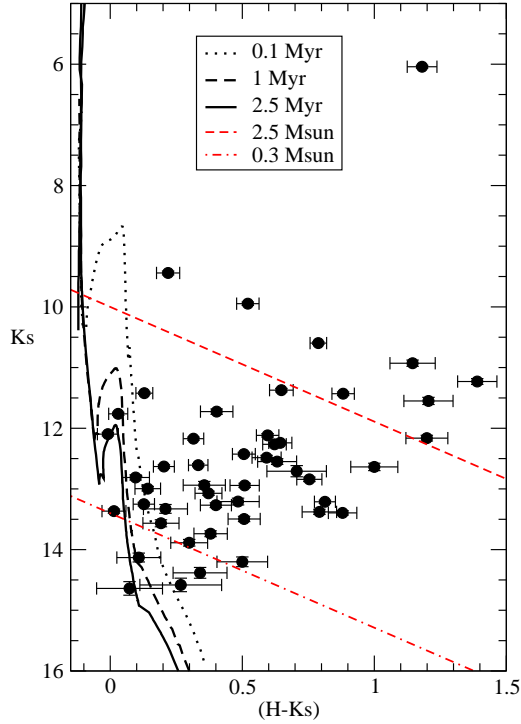}}
\vspace*{-0.2cm}\caption[]{Present $K_s\times(H-K_s)$ CMD of Treasure Chest decontamined data. The PARSEC isochrones are only vertically displaced, likewise \citet{Smith05}. The red lines represent the reddening vectors for stars with masses 2.5 and 0.3 M$_\odot$.}
\label{f12}
\end{figure}

Fig.~\ref{f12} shows a $K_s\times(H-K_s)$ CMD of the Treasure Chest, in order to compare \citet{Smith05} decontamined data with ours. In \citeauthor{Smith05}, the infrared images were taken using SOFI, the near-IR imager of New Technology Telescope (ESO), with $JHK_s$ filters. So, their data achieve deeper magnitudes (completeness limits of $J=19.3$, $H=18.2$ and $K_s=17.2$ mag) than 2MASS data (limits of $J=15.8$, $H=15.1$ and $K_s=14.3$ mag).

\citet{Smith05} present a $K_s\times(H-K_s)$ CMD in their Fig. 8. Comparing the distributions of stars inside the regions bound by the red dashed lines, the number of stars is approximately the same, about 35 to 40 stars. So, despite the fact that their CMDs present some stars with colours as red as $(H-K)\approx$ 2.0-2.5 mag (in this region, our colour $H-K_s$ is limited in $1.4$ mag) and magnitudes $K_s$ up to $18$ mag, the field-star decontamination of both methods present comparable results.

\subsection{Young stellar objects (YSOs)}
\label{sec:3.3}

\begin{figure}
\centering
\vspace{0.1cm}
\resizebox{0.40\textwidth}{!}{\includegraphics{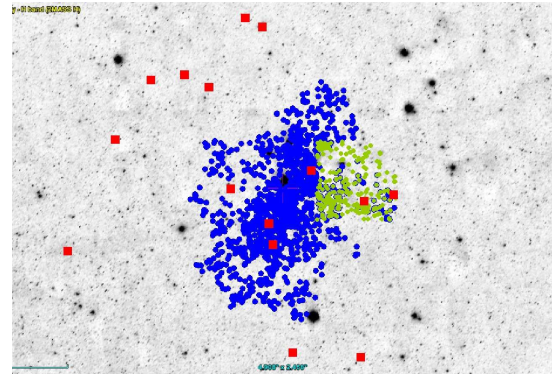}}
\vspace*{-0.2cm}\caption[]{Comparison of the fields of Povich (blue) and Kumar (green) catalogues with our cluster sample, in red ($2^{o}.5\times 3^{o}.5$).}
\label{f13}
\end{figure}

\citet{Povich11} present a catalogue of 1439 new YSOs inside the most active sites of star formation within the Carina Nebula, identified by mid-infrared excess emission. We used this catalogue, which covers a smaller region compared to the present study (Fig.~\ref{f13}), as a reference to verify the PMS stars in common, compared to our clusters sample.

Moreover, we used a catalogue of 467 identified YSOs \citep{Kumar14}, that covers the western part of the Carina nebula. Cross-matching these two catalogues with our clusters, we identified that four of them (objects 1, 7, 12 and 15) are inside the region of \citet{Povich11} and two (objects 3 and 15) are inside the region of \citet{Kumar14}.

\begin{figure}
\centering
\includegraphics[width=0.22\textwidth]{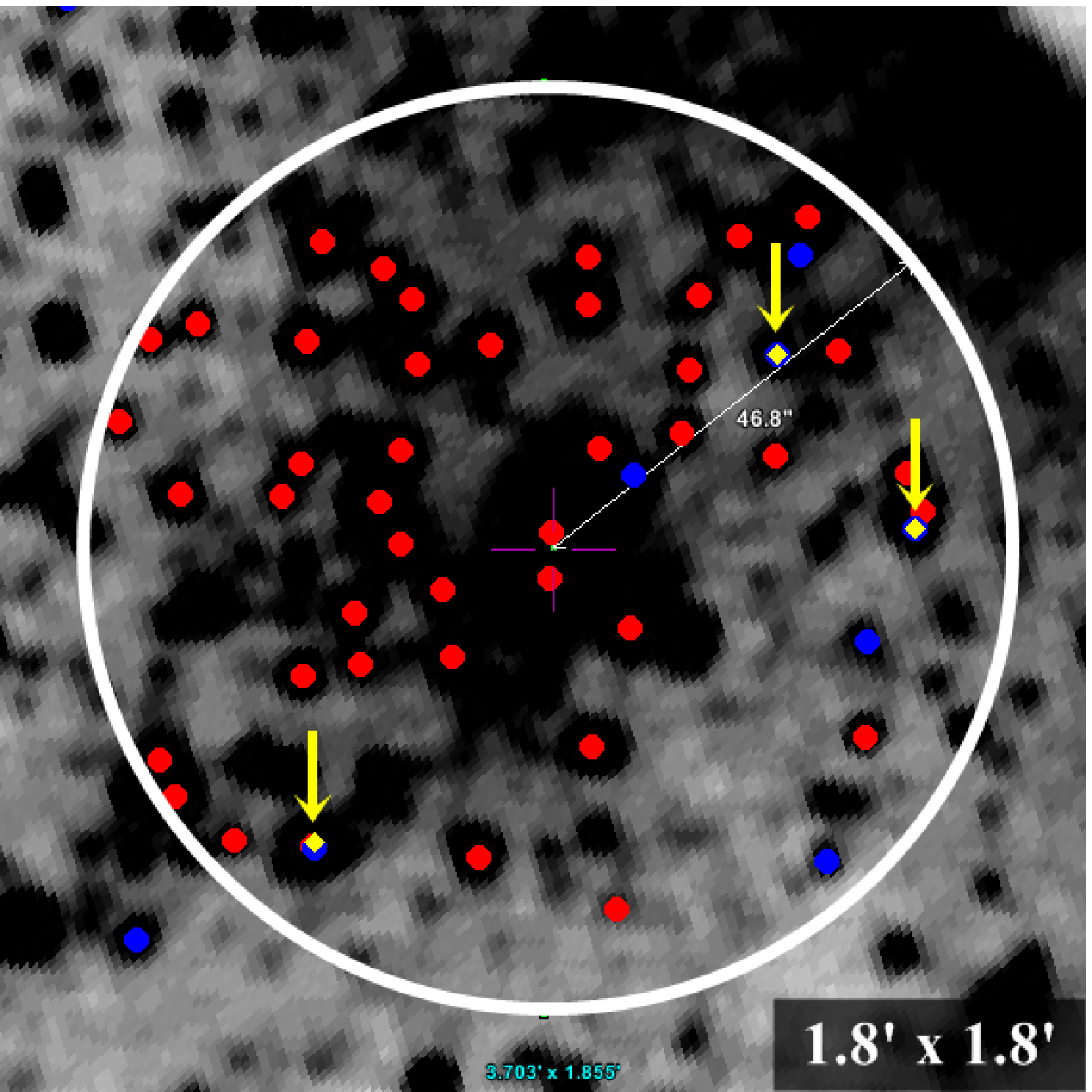}
\hspace{0.3mm}\includegraphics[width=0.22\textwidth]{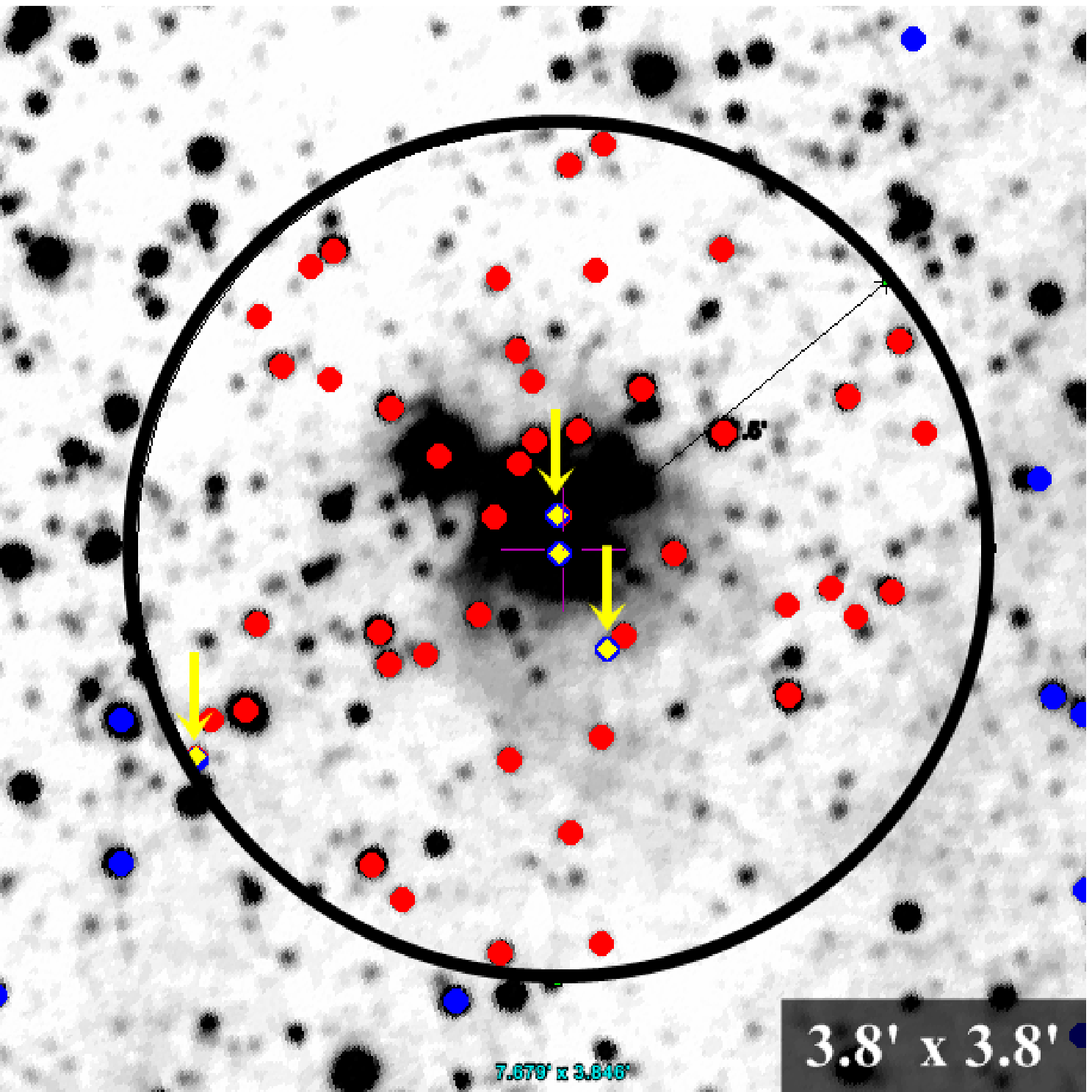}
\vspace*{-0.1cm}\caption[]{Cross-matching of YSOs within the clusters Oliveira\,1 and Oliveira\,12, using the catalogues of \citet{Povich11} and \citet{Kumar14}. Red circles are the decontamined photometry, blue circles are YSOs from the catalogue and the  stars in common are marked in yellow.}
\label{f14}
\end{figure}

The cross-matching is made crossing the coordinates and checking if the magnitudes are comparable. The  YSOs in common (cluster members) are highlighted with a red circle in the respective CMDs and colour-colour diagrams, and their identifiers, coordinates and 2MASS magnitudes are shown in Table~\ref{tab3}. These stars are shown in Fig.~\ref{f14} in yellow, whereas cluster members are marked in red and stars from Povich catalogue are marked with blue circles.

We retrieved 16 YSO stars in common with the two tested catalogues (Table~\ref{tab3}). Considering that our field decontamination approach is statistical, we detect most of the YSOs overlapping the 5 cluster areas (e.g. Fig.~\ref{f14}). Finally, these 5 surveyed clusters have considerably larger intrinsic samples of YSO/PMS stars in the CMDs than the cross-identified YSO stars in the above catalogues.

\begin{table}
\centering
{\footnotesize
\caption{YSOs cross-matching with the clusters members.}
\renewcommand{\tabcolsep}{1.0mm}
\renewcommand{\arraystretch}{1.2}
\begin{tabular}{p{1.33cm}|p{7.1mm}|p{1.35cm}|p{1.4cm}|p{7.3mm}|p{7.3mm}|p{7.3mm}}
\hline
\hline
Cluster&ID&$\alpha(2000)$&$\delta(2000)$&J&H&K$_s$\\
&&(h\,m\,s)&$(^{\circ}\,^{\prime}\,^{\prime\prime})$&$(mag)$&$(mag)$&$(mag)$ \\
($1$)&($2$)&($3$)&($4$)&($5$)&($6$)&($7$)\\
\hline 
Oliveira\,1 &P265\newline P276\newline P299&10:43:36.31\newline 10:43:38.17\newline 10:43:44.43&-59:35:47.9\newline -59:35:30.1\newline -59:36:20.5&15.190\newline 15.521\newline 12.477&13.944\newline14.153\newline11.788&13.015\newline13.262\newline11.277\\
\hline
Oliveira\,3 &K45\newline K84&10:41:00.69\newline 10:40:55.67&-59:49:18.5\newline -59:49:48.2&14.812\newline 14.706&14.086\newline14.046&13.922\newline13.809\\
\hline
Oliveira\,7\newline (CCCP-\newline Cl\,16) &P947\newline P980\newline P997\newline P1014\newline P1016&10:45:39.50\newline 10:45:45.19\newline 10:45:47.47\newline 10:45:50.76\newline 10:45:51.50&-60:06:42.7\newline -60:06:00.8\newline -60:06:08.0\newline -60:04:23.6\newline -60:06:17.3&13.855\newline 13.529\newline 13.234\newline 14.993\newline 14.341&12.869\newline12.516\newline12.777\newline14.004\newline 13.272&12.442\newline12.193\newline12.577\newline13.368\newline12.837\\
\hline
Oliveira\,12\newline (Treasure\newline Chest) &P1019\newline P1025\newline P1026\newline P1074&10:45:52.33\newline 10:45:53.73\newline 10:45:53.74\newline 10:46:03.98&-59:57:24.3\newline -59:57:03.7\newline -59:56:55.7\newline -59:57:47.5&15.364\newline 9.931\newline 13.325\newline 14.768&14.025\newline9.660\newline12.073\newline13.962&13.212\newline9.441\newline10.928\newline13.208\\
\hline
Oliveira\,15 &P1,K4\newline K17&10:39:18.59\newline 10:39:16.94&-59:45:32.6\newline -59:44:18.3&12.417\newline 13.181&12.153\newline12.792&12.169\newline12.520\\
\hline
\end{tabular}
\begin{list}{Table Notes.}
\item Col. $2$: Identifier of each YSO, in the respective catalogue (P refers to \citealp{Povich11}; K refers to \citealp{Kumar14}).  Cols. $3-4$: Equatorial coordinates. Col. $5-7$: 2MASS magnitudes of each YSO.
\end{list}
\label{tab3}
}
\end{table}

\section{Concluding remarks}
\label{sec:4}
In this work we analysed a sample of 101 overdensities, separating 15 of them as best cluster candidates. Three of them were already catalogued (CCCP-Cl\,16, Treasure Chest and FSR\,1555). We derived fundamental parameters for 14 of these candidates projected on the Carina complex, except FSR\,1555 which is an old cluster \citep{Bonatto09}.

The parameters derived for the 12 new candidates confirm that they are, in fact, embedded clusters, with ages ranging from $0.2$ to $4.5$ Myr (which corresponds to the embedded stage of a cluster, \citealp{Lada03}) and distances ranging from 2.5 to 3.4 kpc, comparable with the distance of the Carina complex. The evolutionary stages of the cluster stars were also determined, and we conclude that the great majority of stars belong to the PMS stage, very affected by differential reddening, and the other few stars make up a vertical main-sequence. Most of the new clusters are within shells. This supports the fact that they are extremely young.

The differential reddening values are relatively high, so these clusters are still embedded in their parental cloud of gas and dust. The photometry decontamination was necessary, since the Carina complex is located in a region of low galactic latitude. The automated isochrone fittings were suitable in the CMDs and colour-colour diagrams, although the present ECs are sparsely populated.

We also provided a comparative study between the derived parameters and those given in \citet{Smith05} for the Treasure Chest. The excesses are compatible, the ages and distances are nearly the same, and both data indicate that the member stars are highly embedded. For the common range of $K_s$ magnitudes, \citet{Smith05} and ourselves contain a similar count of members. Our reddest $(H-K_s)$ colours amount to $1.4$ mag. \citet{Smith05} attain colours as red as $(H-K)\approx$ 2.0-2.5 mag. These few stars are below our detection limit in $K_s$. We conclude that the membership in both studies is compatible.

The structure of the clusters, determined by the RDPs, showed that some of them present central concentration of stars, others present a non-central density peak of stars, and the remaining have several density peaks and low stellar counts. We conclude that the existence of a cluster cannot be determined only by a centrally condensed RDP. Aditional CMDs and colour-colour diagrams for the present sample are given in electronic form.

The candidates presenting CMDs and RDPs of clusters, are now classified undoubtedly as low-mass clusters, according to our results. Those with CMDs that meet cluster expectations, but show irregular RDPs can still be considered candidates. We may be dealing in both cases with the long sought lowest mass clusters that will soon dissolve into the field \citep{Lada03}, feeding it with fresh stars. In addition to the present objects having cluster CMD and/or RDP, they are embedded in shells and bubbles as revealed especially by WISE and Herschel, which strongly supports them as star-forming physical systems. Further analysis are welcome, especially to probe lower mass PMS stars, by means of deeper and small pixel size photometry, and with proper motion analyses, using Gaia data wherever possible, owing to possible high absorption limitation effects.

\vspace{0.45cm}

\textit{Acknowledgements}: We thank an anonymous referee for interesting remarks. This publication makes use of data products from the Two Micron All Sky Survey, which is a joint project of the University of Massachusetts and the Infrared Processing and Analysis Centre/California Institute of Technology, funded by the National Aeronautics and Space Administration and the National Science Foundation. E. B. and C. B. acknowledge support from CNPq (Brazil). R. O. acknowledges a fellowship from INCT-A (Brazil).


\label{lastpage}
\end{document}